\documentclass[acmsmall,screen]{acmart}
\AtBeginDocument{%
  }

\setcopyright{acmlicensed}
\copyrightyear{2026}
\acmYear{2026}
\acmDOI{XXXXXXX.XXXXXXX}

\acmJournal{CSUR}
\acmVolume{37}
\acmNumber{4}
\acmArticle{111}
\acmMonth{8}

\usepackage{amsmath,amsfonts}
\usepackage{algorithmic}
\usepackage{algorithm}
\usepackage{array}
\usepackage{textcomp}
\usepackage{stfloats}
\usepackage{url}
\usepackage{verbatim}
\usepackage{graphicx}
\usepackage[table]{xcolor}
\usepackage{multirow}
\usepackage{acronym}
\usepackage{pifont} 
\usepackage{subcaption}
\usepackage{xurl}
\usepackage{makecell}
\usepackage{booktabs}
\usepackage{hyperref}
\usepackage{cleveref}
\usepackage{enumitem}

\usepackage{lineno}
\usepackage{ulem}

\Crefname{appendix}{Appendix}{Appendices}

\begin{document}

\acrodef{SM}[SM]{Streaming Multiprocessor}
\acrodef{PB}[PB]{Processing Block}
\acrodef{TDP}[TDP]{Thermal Design Power}
\acrodef{MAC}[MAC]{Multiply–Accumulate}
\acrodef{INT}[INT]{Integer}
\acrodef{INT4}[INT4]{4-bit Integer}
\acrodef{INT8}[INT8]{8-bit Integer}
\acrodef{INT32}[INT32]{32-bit Integer}
\acrodef{FP}[FP]{Floating Point}
\acrodef{FP8}[FP8]{8-bit Floating Point}
\acrodef{FP16}[FP16]{16-bit Floating Point}
\acrodef{FP32}[FP32]{32-bit Floating Point}
\acrodef{FP64}[FP64]{64-bit Floating Point}
\acrodef{BF16}[BF16]{16-bit Brain Floating Point}
\acrodef{TF32}[TF32]{TensorFloat-32}
\acrodef{GPU}[GPU]{Graphics Processing Unit}
\acrodef{CUDA}[CUDA]{Compute Unified Device Architecture}
\acrodef{RT}[RT]{Ray Tracing}
\acrodef{AI}[AI]{Artificial Intelligence}
\acrodef{TFLOPS}[TFLOPS]{Tera Floating Point Operations per Second}
\acrodef{TOPS}[TOPS]{Tera Operations per Second}
\acrodef{GFLOPS}[GFLOPS]{Giga Floating Point Operations per Second}
\acrodef{CAGR}[CAGR]{Compound Annual Growth Rate}
\acrodef{HBM}[HBM]{High Bandwidth Memory}
\acrodef{DDR}[DDR]{Double Data Rate}
\acrodef{DRAM}[DRAM]{Dynamic Random Access Memory}
\acrodef{GDDR}[GDDR]{Graphics Double Data Rate}
\acrodef{GB}[GB]{Gigabytes}
\acrodef{GBS}[GB/s]{Gigabytes per second}
\acrodef{PPT}[ppt]{percentage point}
\acrodef{MIMD}[MIMD]{Multiple Instruction Multiple Data}
\acrodef{SIMD}[SIMD]{Single Instruction Multiple Data}
\acrodef{SIMT}[SIMT]{Single Instruction Multiple Thread}
\acrodef{PCIe}[PCIe]{Peripheral Component Interconnect Express}
\acrodef{SXM}[SXM]{Server PCI Express Module}
\acrodef{CPIU}[CPI-U]{Consumer Price Index for All Urban Consumers}
\acrodef{BIS}[BIS]{Bureau of Industry and Security}
\acrodef{TPP}[TPP]{Total Processing Performance}
\acrodef{IC}[IC]{Integrated Circuit}
\acrodef{IO}[I/O]{Input/Output}
\acrodef{ECCN}[ECCN]{Export Control Classification Number}
\acrodef{AGR}[AGR]{Annual Growth Rate}
\acrodef{DT}[DT]{Doubling Time}
\acrodef{PLX}[PLX]{Peripheral Component Interconnect Express Switch}
\acrodef{NV-HBI}[NV-HBI]{NVIDIA High-Bandwidth Interface}
\acrodef{NPU}[NPU]{Neural Processing Unit}
\acrodef{ML}[ML]{Machine Learning}
\acrodef{DPU}[DPU]{Data Processing Unit}
\acrodef{CPU}[CPU]{Central Processing Unit}
\acrodef{CU}[CU]{Compute Unit}
\acrodef{OCP}[OCP]{Open Compute Project}
\acrodef{MAD}[MAD]{Multiply–Add}
\acrodef{FMA}[FMA]{Fused Multiply–Add}
\acrodef{NVFP4}[NVFP4]{NVIDIA 4-bit Floating Point}
\acrodef{MXFP6}[MXFP6]{Microscaling 6-bit Floating Point}
\acrodef{MXFP4}[MXFP4]{Microscaling 4-bit Floating Point}
\acrodef{ALU}[ALU]{Arithmetic Logic Unit}

\newcommand\newtext[1]{\textcolor{blue}{#1}}
\newcommand\newnewtext[1]{\textcolor{brown}{#1}}
\newcommand\ken[1]{\textcolor{green}{#1}}
\newcommand\myworries[1]{\textcolor{red}{#1}}
\newcommand\partialdata[1]{\textcolor{orange}{#1}}
\newcommand\secref[1]{\S~\ref{#1}}
\newcommand{\ourparagraph}[1]{\textbf{#1:}}

\newcommand{\cmark}{\checkmark}%
\newcommand{\xmark}{\ding{53}}%
\newcommand{\myna}{$-$}%
\newcommand{\myh}{$H$}%
\newcommand{\myhalfcheck}{\sout{\cmark}}

\newcommand{\gna}{\textsuperscript{$\dagger$}}
\newcommand{\gnaa}{\textsuperscript{$\diamondsuit$}}
\newcommand{\gne}{\textsuperscript{$\wr$}}
\newcommand{\gno}{\textsuperscript{$\star$}}
\newcommand{\gni}{\textsuperscript{$\bullet$}}
\newcommand{\gnu}{\textsuperscript{$\odot$}}
\newcommand{\gnaao}{\textsuperscript{$\spadesuit$}}
\newcommand{\gnuu}{\textsuperscript{$\clubsuit$}}
\newcommand{\gnuua}{\textsuperscript{$\heartsuit$}}
\newcommand{\gnuuo}{\textsuperscript{$\triangle$}}

\title{How Much Progress Has There Been in NVIDIA Datacenter GPUs?}




\author{Emanuele Del Sozzo}
\email{delsozzo@mit.edu}
\orcid{0000-0003-3101-8118}
\affiliation{%
  \institution{MIT FutureTech, Computer Science and Artificial Intelligence Laboratory (CSAIL), Massachusetts Institute of Technology}
  \city{Cambridge}
  \state{MA}
  \country{USA}
}

\author{Martin Fleming}
\email{marti264@mit.edu}
\orcid{0009-0009-0413-1909}
\affiliation{%
  \institution{MIT FutureTech, Computer Science and Artificial Intelligence Laboratory (CSAIL), Massachusetts Institute of Technology}
  \city{Cambridge}
  \state{MA}
  \country{USA}
}

\author{Kenneth Flamm}
\email{kflamm@mit.edu}
\orcid{0000-0003-4217-2137}
\affiliation{%
  \institution{MIT FutureTech, Computer Science and Artificial Intelligence Laboratory (CSAIL), Massachusetts Institute of Technology}
  \city{Cambridge}
  \state{MA}
  \country{USA}
}

\author{Neil Thompson}
\email{neil\_t@mit.edu}
\orcid{0000-0001-9888-573X}
\affiliation{%
  \institution{MIT FutureTech, Computer Science and Artificial Intelligence Laboratory (CSAIL), Massachusetts Institute of Technology}
  \city{Cambridge}
  \state{MA}
  \country{USA}
}

\renewcommand{\shortauthors}{Del Sozzo et al.}

\begin{abstract}
As the role of modern \acp{GPU} becomes increasingly essential for several computing tasks, analyzing their past and current progress is paramount for determining future constraints on scientific research. This is particularly compelling in the \ac{AI} domain, where rapid technological advancements and fierce global competition have led the United States to recently implement export control regulations limiting international access to advanced AI chips. Consequently, this paper examines technical progress in NVIDIA datacenter GPUs from the mid-2000s through 2025. Our main results identify doubling times of 1.43 and 1.67 years for FP16 and FP32 dense operations, while FP64 doubling times range from 2.05 to 3.79 years. 
Off-chip memory size and bandwidth have grown at slower rates than computing performance, doubling every 3.29 to 3.41 years, whereas the release prices and power consumption roughly doubled every 5.03 and 15 years, respectively.
Moreover, our cross-vendor comparison of the top-performing GPUs per year shows that NVIDIA’s performance advantage is narrowing, but not enough to compel a major market shift. 
Finally, we quantify the potential implications of current U.S. export control regulations and the consequent performance gaps, which the recently proposed policy changes could shrink from 23.6$\times$ to 3.54$\times$.
\end{abstract}





\maketitle

\section{Introduction}\label{sec:introduction}

The growing need for computational power dedicated to graphics rendering motivated the the creation of the specialized hardware we commonly call \acfp{GPU}. 
Initially designed for to 2D/3D graphics and gaming, \acp{GPU} became a performant architecture to accelerate highly parallel workloads. 
According to the June 2025 TOP500 list~\cite{top500}, all five of the world's top supercomputers employ GPUs. 
Moreover, GPUs have also become the \textit{de facto} standard for the training and inference of \acf{AI} models, transforming the computer science field and downstream industrial markets~\cite{gpu_science, gpu_evolution}.
Indeed, as the central core of \ac{AI} operations can be reduced to matrix/vector multiplications, the massively parallel architecture of GPUs makes them more suitable and efficient for such computations than \acp{CPU}.
The role of GPUs (and AI chips in general) has become so critical in the current intense global competition over AI development that U.S. national security policy has introduced export control regulations, intended to restrict access to advanced AI chips by certain countries~\cite{ic_bis_2022, ic_bis_2023, hbm_bis_1, hbm_bis_2, ic_bis}.

GPU hardware improvements have been a key factor in the current AI revolution, alongside improved algorithms and larger datasets~\cite{9623445}. 
As the characteristics and requirements of large-scale AI models increase at impressive rates (e.g., computing demand doubles every 6 to 10 months~\cite{epoch2025trainingcomputedecomposition, Sevilla_2022} and the longest context windows grow by 30$\times$ per year~\cite{epoch2025contextwindows}), the GPU landscape has changed substantially at both the software and hardware levels to keep sustaining AI progress.
Over the years, GPU vendor NVIDIA released highly optimized AI software libraries (e.g., CuDNN and Transformer Engine) within its \ac{CUDA} framework. 
NVIDIA also integrated AI-oriented hardware functionality inside its GPUs, such as \ac{FP16} units and Tensor Cores, alongside standard compute units for \ac{FP32} and \ac{FP64}. 
The improved performance is now measured in \ac{TFLOPS}~\cite{p100}. 
These innovations demonstrate how domain specialization became a viable pathway, at least temporarily, to continued performance improvement~\cite{4785860, moore1975progress, 7878935, leiserson2020there} and performance “scaling laws” (e.g., Huang’s law for AI~\cite{8352557}).
In contrast to the empirical record for CPU performance during the heyday of Moore’s Law and Dennard scaling in the late twentieth century~\cite{flamm2021measuring}, performance improvements and new computing features for GPUs have generally come at the cost of greater energy use and higher prices for modern GPUs. 
For instance, if we compare top server-/datacenter-class CPUs and GPUs from 2025 (e.g., Intel Xeon 6978P~\cite{intel_xeon_6978p}, AMD Ryzen Threadripper PRO 9995WX~\cite{amd_ryzen_9995wx}, and Blackwell Ultra B300~\cite{blackwell_ultra}), the NVIDIA device is 2.2$\times$ to 3.1$\times$ more power hungry and almost 5$\times$ more expensive than Intel and AMD devices.\footnote{This comparison uses \ac{TDP} as a power-consumption metric. We respectively extrapolate and approximate the TDP and release price of a single B300 GPU from the available information for its HGX version.}
These continuous innovations at the software and hardware levels, along with the exploding need for more (and more powerful) GPUs, made NVIDIA the \textit{de facto} leader in this market, with a market capitalization of \$4.5 trillion in early 2026~\cite{nvidia_capitalization}.
In general, GPU users want fast turnaround times for AI training but also high precision (for instance, for scientific computing); at the same time, cutting-edge AI models are typically growing larger with billions of parameters. Thus, analyzing the evolution of GPU microarchitectures is fundamental to finding the right balance among these conflicting demands and identifying the trade-offs NVIDIA chose to address customers' needs.

\begin{figure*}
    \centering
    \includegraphics[width=\textwidth]{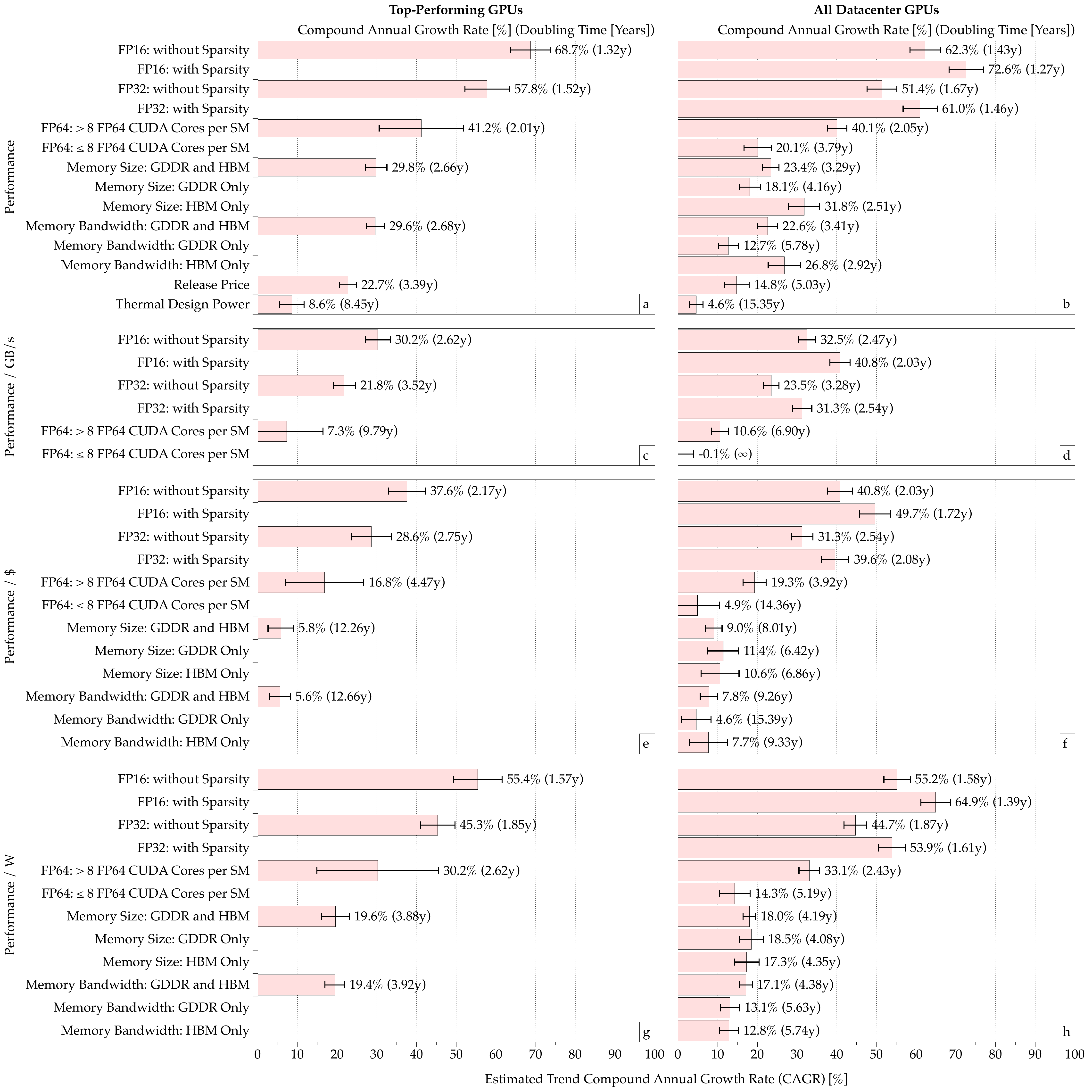}
    \caption{\acfp{CAGR} with 90 percent confidence intervals and \acfp{DT} for computing performance, namely \acf{FP16}, \acf{FP32}, and \acf{FP64}, off-chip memory size and bandwidth, release price, and \acf{TDP} deriving from technical improvements (subplots \textit{a} and \textit{b}) and their per-memory bandwidth (\textit{c} and \textit{d}), per-dollar (\textit{e} and \textit{f}), and per-watt (\textit{g} and \textit{h}) ratios for top-performing GPUs (left column) and all datacenter GPUs (right column).
    The Figure also accounts for the impact of sparsity (\ac{FP16} and \ac{FP32}) and differentiates between \acp{GPU} with a high/low number of \ac{FP64} \ac{CUDA} Cores per \acf{SM}. 
    Similarly, it shows the separate and joint contributions of GDDR and HBM technologies on memory size and bandwidth.}
    \label{fig:cagr_histogram}
\end{figure*}

Given the importance of GPU technical trends in current scientific fields and policy debates, this paper analyzes the pace of technical progress in NVIDIA datacenter GPUs released from the mid-2000s through 2025 and measures key trends across relevant metrics, including computing performance, memory size and bandwidth, power consumption, and release price. 
\Cref{fig:cagr_histogram} shows a summary of our results in terms of \acfp{CAGR} and \acfp{DT} for both a subset of the top-performing NVIDIA GPUs per year and all NVIDIA datacenter GPUs.
Previous literature has examined the progress and trends of GPUs and AI hardware from various perspectives; however, these studies have important limitations.
For instance, Huang’s law~\cite{8352557, huang_law} focuses on the peak computing performance across a narrow and recent set of NVIDIA GPU microarchitectures and mixes the level of precision (FP16 down to FP4), despite the reduced functionality of the smaller bitwidth systems. Not only is it implausible that such bitwidth improvements could continue, but it also leads to an inflated GPU progress estimate (nearly double what we find). 
Conversely, other studies focus on the Machine Learning (ML) and AI domains~\cite{epoch2023trendsinmachinelearninghardware, EpochMachineLearningHardware2024}, or on the impact of generative AI on productivity~\cite{baily2025generative,baily2026generative},
and explore a broad range of devices across vendors (e.g., NVIDIA and AMD), types (e.g., GPUs and specialized hardware), and classes (e.g., desktop, workstation, and datacenter devices). These studies also have important drawbacks. First, their analysis gives a great deal of weight to seldom-used systems, rather than the dominant platforms. They also fail to group AI-oriented (tensor) operations with others~\cite{epoch2023trendsinmachinelearninghardware, EpochMachineLearningHardware2024}, which we argue should be grouped to correctly characterize performance improvements,
or miss the contribution of such operations totally~\cite{baily2025generative,baily2026generative}. As a consequence of these choices, some analyses find FP16 progress rates that are 26 to 37 \acp{PPT} lower than ours~\cite{epoch2023trendsinmachinelearninghardware, EpochMachineLearningHardware2024}.
By focusing on NVIDIA datacenter GPUs, by far the most widely used chips for AI workloads, we can simultaneously measure technical trends in an apples-to-apples comparison and provide a practical view of the rates of progress faced by most firms using AI.

Our analysis identifies the following key insights:
\begin{enumerate}[label=Insight \arabic* -,leftmargin=*]
    \item GPU progress surpassed Moore's Law to keep pace with AI growth;
	\item chip-level enhancements boosted GPU performance;
	\item the computing power of FP16 and FP32 doubled in less than 2 years;
	\item FP64 compute units became a lower-priority resource;
	\item HBM boosted off-chip memory size and bandwidth;
	\item the growth in release price and TDP for best-in-class GPUs is almost 2$\times$ that of the average for the broad datacenter lineup;
	\item computing power grew much faster than off-chip memory bandwidth;
	\item technical improvement per watt grew faster than improvement per dollar;
    \item AMD and Intel GPUs narrowed the gap, but not enough to significantly affect NVIDIA’s dominance;
	\item the recent update to export control regulations reduced the controlled-to-uncontrolled performance gap from 23.6$\times$ to 3.54$\times$.
\end{enumerate}


In the remainder of this paper, we first provide a simplified architectural review of modern NVIDIA GPUs (\secref{sec:background_gpu}). 
We then report our data collection, selection, and analysis methodology (\secref{sec:data_and_method}). 
Next, we present our analysis of rates of improvement for NVIDIA GPUs, focusing first on the top-performing GPUs per year, and then on all the datacenter GPUs in our dataset (\secref{sec:gpu_progress_analysis}). 
We also compare the top-performing datacenter GPUs from NVIDIA, AMD, and Intel, year by year (\secref{sec:vendor_comparison}).
Then we examine the United States export control regulations and estimate the potential effects of these restrictions on the affected countries (\secref{sec:export}). We compare our work with the current literature and discuss both the limitations of our analyses and potential future research directions (\secref{sec:limitations}). 
Finally, we conclude by summarizing our results (\secref{sec:conclusion}).


\section{NVIDIA Datacenter GPU Architecture}\label{sec:background_gpu}

\begin{figure*}[t]
\centering
\includegraphics[width=0.92\textwidth]{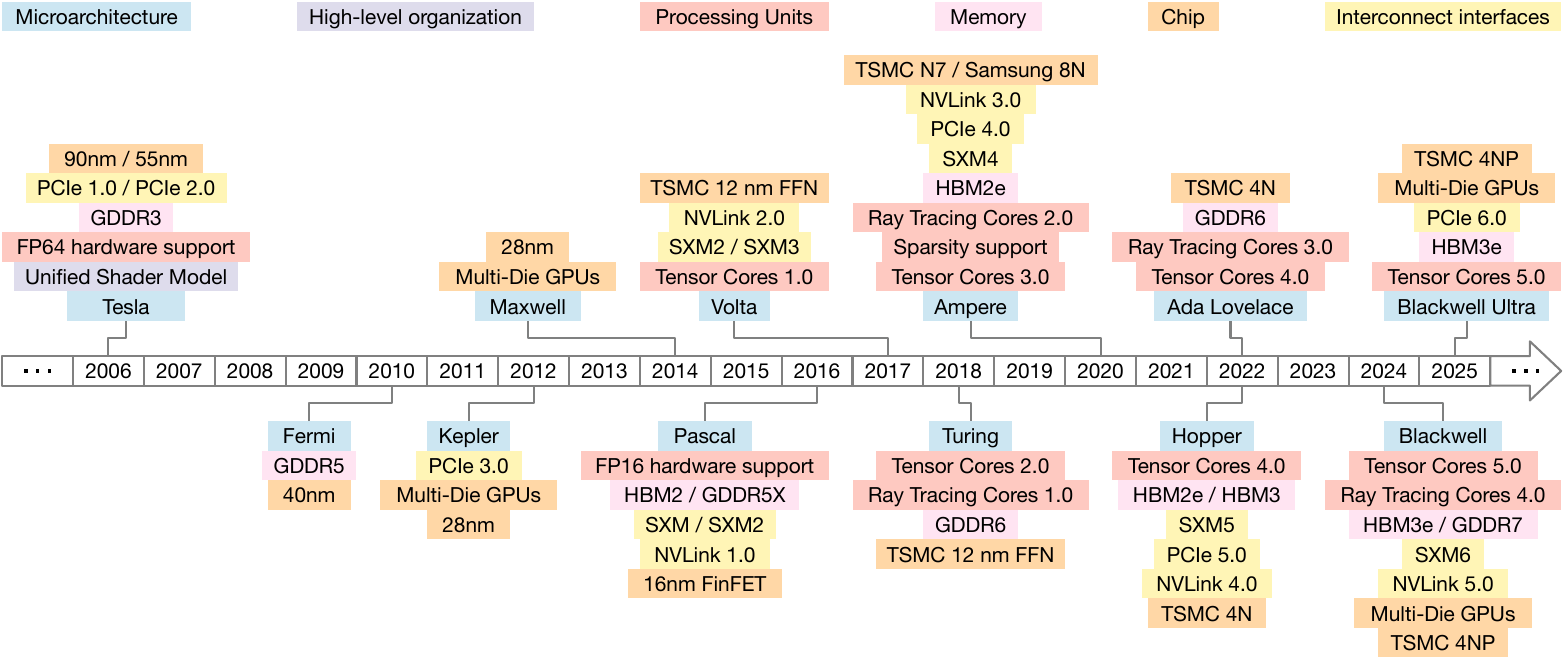} 
    \caption{Timeline of NVIDIA microarchitectures from 2006. For each microarchitecture, we report the main architectural features and novelties that characterized it throughout its lifetime (not necessarily available at the release year) and are relevant to our analysis of NVIDIA datacenter GPUs.}
    \label{fig:timeline}
\end{figure*}

This section provides a historical overview of the main architectural characteristics of NVIDIA datacenter GPUs relevant to this paper and implemented since 2006, the year the Tesla microarchitecture was released, the first to feature datacenter GPUs; therefore, other features, such as the \acf{CUDA} framework and its libraries, which are generally as important as those discussed in this section and the following ones, are out of scope for this paper.
This background section is intended primarily for readers unfamiliar with NVIDIA GPU architecture to facilitate understanding of our paper. For this reason, we organize this section into different parts that target relevant topics in NVIDIA datacenter GPU architectures. The timeline in \Cref{fig:timeline} follows the same approach to illustrate the introduction of key features over the years.

\textit{High-Level Organization:} Starting with the Tesla microarchitecture (2006), NVIDIA adopted the \textit{unified shader model}, which turned the separate functional units (e.g., pixel and vertex shaders) implemented in previous microarchitectures into a homogeneous set of processing cores, grouped into \acfp{SM}, capable of performing a broader range of operations. As microarchitectures changed over time, these cores and their functionalities became more heterogeneous (see later for more details), 
the number and types of cores per SM increased, and GPU architectures generally evolved to support not only graphics rendering but also scientific computing.

Currently, modern NVIDIA \acp{GPU} are computing devices that implement a two-level compute hierarchy: a parallel collection of \acp{SM} running in \ac{MIMD} mode, and \acp{SM} consisting of several heterogeneous processing cores running in \ac{SIMT} mode.
These \ac{SIMT} compute engines may be viewed as a more flexible descendant of the historical \ac{SIMD} engine model used in earlier parallel computers and coprocessors, as they schedule processing resources to execute an identical instruction on a set of multiple data processing threads every clock cycle. 
The incorporation of multiple parallel compute cores into NVIDIA \acp{GPU} roughly coincided with the historical adoption of multi-core processors in general-purpose computer processor designs.
The \ac{SM} compute cores can also be partitioned into separate ``virtual'' \ac{MIMD} engines 
to increase programming flexibility or enable multiple users to access resources on a single GPU.

\textit{Processing Cores:} Each \ac{SM} provides hardware support for multiple instances of an instruction being executed simultaneously, with the number of configurable simultaneous \ac{INT} or \ac{FP} operations dependent on the arithmetic precision, which may range from low-bit precision (e.g., 16 bits) to double precision (64 bits) to handle operations requiring highly accurate results. 
The math processing cores are called Streaming Processors or \ac{CUDA} Cores, and an \ac{SM} contains variable amounts and types of \ac{CUDA} Cores depending on the \ac{GPU} microarchitecture and model. 
For instance, \acp{SM} initially featured separate or shared \ac{CUDA} Cores, permitting only 32-bit integer and \ac{FP32} operations, while FP64 hardware support was introduced in 2008 with the release of the Tesla GT200 chip. These cores are \ac{MAC} units, as they can perform two arithmetic operations (multiplication and addition/accumulation) per clock cycle. Initially, CUDA Cores in the Tesla microarchitecture implemented \ac{MAC} through the \ac{MAD} instruction; then, in the Fermi microarchitecture (2010), NVIDIA switched to the \ac{FMA} instruction, which requires a single rounding step rather than two (when executed on \ac{FP} numbers), without losing precision in the addition. In 2016, with the Pascal microarchitecture, NVIDIA introduced hardware support for \ac{FP16} within \ac{CUDA} Cores, enabling more performant execution of AI applications, which previously could use 16-bit precision only through software emulation. Generally, \ac{FP16} calculations are faster and less area-demanding than \ac{FP32} operations and also utilize half the memory.

Starting with the Volta microarchitecture (2017) and subsequent ones, NVIDIA introduced specialized Tensor Cores alongside CUDA Cores within the \acp{SM} to accelerate tensor computations, which are frequently used in the AI domain, through multiple \ac{MAC} operations per clock cycle.\footnote{Tensor computation refers to the mathematical operations and manipulations performed on tensors, namely multidimensional arrays of data.} 
For instance, the first generation of Tensor Cores multiplied two 4$\times$4 \ac{FP16} matrices and then added a third \ac{FP16} (or \ac{FP32}) matrix, producing a new \ac{FP16} (or \ac{FP32}) matrix through 64 \ac{MAC} operations per clock cycle. 
The following generations of Tensor Cores extended support to larger input matrices, which implies more MACs per clock cycle, and additional data precisions beyond \ac{FP16}, including \ac{INT8}, \ac{BF16} (also supported by recent CUDA Cores), \ac{TF32}, and \ac{FP64}. Moreover, the fifth generation (2024) adds native support for sub-8-bit data types, such as the \ac{OCP} \ac{MXFP6} and \ac{MXFP4} formats~\cite{rouhani2023microscaling, mx_specification}, as well as the \ac{NVFP4}~\cite{nvfp4}.
On the other hand, starting from the third generation (2020), Tensor Cores implement hardware support for sparse matrices at specific data precisions, which can significantly accelerate computations in certain AI applications.\footnote{Data sparsity denotes the condition in which a large proportion of data in a dataset is zero, null, or missing. The structured sparsity supported by NVIDIA GPUs requires a 2:4 sparsity pattern, where at least 2 values in each contiguous group of 4 must be zero (50\% sparsity rate).}

Finally, although they are outside the scope of this paper, it is worth noting that \acp{SM} include other units/cores, such as texture units to accelerate storage and retrieval of pixel characteristics in images or simulations, and \ac{RT} Cores (in desktop, workstation, and some datacenter \acp{GPU}) to accelerate the calculation of pixel characteristics in images or simulations.

\textit{Memory:} 
NVIDIA GPUs feature different types of on-chip and off-chip memory. 
Each \ac{SM} includes internal registers, a specific number assigned to each scheduled thread, a Level zero (L0) instruction cache, and a combined Level one (L1) data cache and shared memory, enabling low-latency, high-bandwidth data exchange between threads. Moreover, the Hopper microarchitecture (2022) introduced the Tensor Memory Accelerator (TMA), which supports bidirectional, asynchronous transfers between shared and global memory~\cite{h100}. 
On the other hand, NVIDIA GPUs also incorporate Level two (L2) data caches that reside outside the SM hierarchy.

Each \ac{SM} can also access off-chip memory through on-chip memory controller units. Off-chip memory \ac{DRAM} can either have a very wide interface and very low latency, provided by High Bandwidth Memory (HBMx), or less performant but less expensive solutions, such as Graphics Double Data Rate (GDDRx) memory, where the trailing ``x'' is the version of the industry standard. 
Starting with the Pascal microarchitecture (2016), NVIDIA began to employ HBM technology for the off-chip memories of its datacenter GPUs, while desktop/workstation GPUs still rely on GDDR-based memories (e.g., the Turing and Ada Lovelace microarchitectures, as well as some Ampere chips).
The most current and fastest version of \acsu{HBM} is HBM3e, 
while the most advanced \acsu{GDDR} memory in volume production is GDDR7, introduced in 2024~\cite{gddr_evolution}.

In general, off-chip memories exhibit various features that determine their performance and suitability for specific tasks. In this paper, we focus on the off-chip memory size/capacity and bandwidth; of course, other characteristics, such as the latency, are equally relevant.
The former indicates the number of bytes the off-chip memory can store; the latter indicates how quickly bytes can be transferred between the off-chip memory and the computing chip. The memory interface width and standard used determine the theoretical off-chip memory bandwidth. These features are particularly relevant in applications whose performance is bounded by off-chip memory capabilities, as such applications require storing large amounts of data in off-chip memory and frequently transferring data to/from the computing chip.

\textit{Chip:} As chip manufacturing improved over the years, NVIDIA GPUs were designed to use various newly introduced technology nodes. While NVIDIA employed standard technology nodes for its initial datacenter GPU chips (e.g., 90nm), recent devices leverage custom manufacturing processes specifically tailored to NVIDIA GPUs, such as TSMC 4NP. On the other hand, as the size of GPU chips continued to grow, often approaching the limits of current chip fabrication technology (constrained fundamentally by the size of the reticle/mask stepped by optical patterning equipment across the surface of a wafer in order to fabricate individual chips on a silicon wafer), NVIDIA decided to move from a single-die (or single-chip) to a multi-die approach for its Blackwell microarchitecture (2024) and subsequent microarchitectures~\cite{gtc_2025}. 
Actually, this is not the first time that NVIDIA has gone down this multi-die path for its GPUs, as some Kepler (2012) and Maxwell (2015) boards (i.e., K10, K80, M10, and M60) already featured multi-die GPUs. Use of multi-die GPUs can also deliver economic benefits, since smaller dies generally deliver higher yields of non-defective chips during the manufacturing process.
However, while dies on older GPUs, such as the K80, were connected via a relatively slow on-board \ac{PLX}~\cite{k80_overview}, modern GPUs rely on custom, power-efficient die-to-die interfaces, namely the \ac{NV-HBI}~\cite{blackwell_ultra}.

NVIDIA also developed ``superchips'' that combine a Grace CPU (licensing the ARMv9.0 instruction set architecture) with one or two GPUs~\cite{grace_hopper,grace_blackwell} and can potentially operate as standalone systems (with NVIDIA's computing software stacks running on the Grace CPU) or in combination with other superchips. Similarly, a few NVIDIA devices combine both a GPU and a \ac{DPU}~\cite{bluefield} on the same board, resulting in a converged acceleration card for computing and networking~\cite{converged_accelerators_1, converged_accelerators_2}.

\textit{Interconnect Interfaces:} 
Each GPU (and its SMs) can communicate with other GPUs and external memory storage devices through on-board interconnect interfaces: NVLink, \ac{PCIe}, or \ac{SXM}.
The external interface speed affects overall computation performance in clusters of NVIDIA datacenter \acp{GPU}. 
Although a thorough analysis of characteristics of communication with external resources (e.g., other computing or memory devices) is beyond the scope of this paper, we will examine the critical role of the interconnection bandwidth in export control regulations (\secref{sec:export}).

\section{Data Collection, Selection, and Methodology}\label{sec:data_and_method}

This section describes our data collection procedures for the class of NVIDIA GPUs we target in this work (\secref{subsec:data_collection}), explains the selection process for the metrics of interest and the top-performing GPUs per year (\secref{subsec:data_selection}), and presents the analysis methodology we employed on our curated GPU dataset to derive progress trends (\secref{subsec:methodology}).


\begin{table*}[t]
    \caption{Collected datacenter \acp{GPU} in our dataset}
    \label{table:collected_gpus}
    \begin{minipage}{\textwidth}
    \centering
    \resizebox{\textwidth}{!}{
    \begin{tabular}{ccl}
    \toprule
        \multicolumn{2}{c}{\textbf{Microarchitecture}} & \textbf{GPU} \\
         \textbf{Name} & \textbf{First Release} & \textbf{Model}\textsuperscript{$\dagger$} \\
        \midrule
        Tesla & 2006 & C870\textsuperscript{*}, C1060\textsuperscript{*}, C1080, M1060\textsuperscript{*} \\
        \addlinespace
        Fermi & 2010 & C2050, C2070\textsuperscript{*}, C2075, C2090, M2050, M2070, M2070Q, M2075, M2090, X2070, X2090 \\
        \addlinespace
        Kepler & 2012 & K8, K10\textsuperscript{*}, K20c, K20m, K20s, K20X, K20Xm, K40c\textsuperscript{*}, K40d, K40m, K40s, K40st, K40t, K80\textsuperscript{*} \\
        \addlinespace
        Maxwell & 2014 & M4, M10, M40, M60\textsuperscript{*} \\
        \addlinespace
        Pascal & 2016 & P4, P10, P40, P100\textsuperscript{*}, P100 DGXS \\
        \addlinespace
        Volta & 2017 & PG500-216, PG503-216, V100\textsuperscript{*}, V100 DGXS, V100 FHHL, V100S\textsuperscript{*} \\
        \addlinespace
        Turing & 2018 & T10, T4, T4G, T40 \\
        \addlinespace
        \multirow{2}{*}{Ampere} & \multirow{2}{*}{2020} & A2, A10, A10G, A10M, A16, A30, A30X, A40, A100\textsuperscript{*}, A100X\textsuperscript{*}, A800, AX800, PG506-207, PG506-217,\\
        & & PG506-232, PG506-242 \\
        \addlinespace
        Ada Lovelace & 2022 & L2, L4, L20, L40, L40G, L40S, L40 CNX\\
        \addlinespace
        Hopper & 2022 & H20, H100\textsuperscript{*}, H100 CNX, H100 NVL\textsuperscript{*}, H200, H200 NVL, H800 \\
        \addlinespace
        Blackwell & 2024 & B100, B200\textsuperscript{*}, RTX PRO 6000 Server \\
        \addlinespace
        Blackwell Ultra & 2025 & B300\textsuperscript{*} \\
        \bottomrule
    \end{tabular}
    }
\footnotesize{\textsuperscript{$\dagger$}For simplicity, this list excludes GPU model variants sharing the same name but implementing different features (e.g., off-chip memory size or form factor)}.\\
\footnotesize{\textsuperscript{*}This GPU model (or its variants) is part of the top-performing GPUs per year subset.}
\end{minipage}
\end{table*}

\subsection{Data Collection}\label{subsec:data_collection}

Our work focuses on NVIDIA, dominant in both desktop and datacenter GPU sales, with over 80\% and 90\% market shares, respectively~\cite{jonpeddie,bloomberg}. 
NVIDIA offers a broad range of GPUs with every microarchitecture covering multiple computing domains. For instance, considering only GPUs based on the Ampere microarchitecture (released in 2020 and accounting for more than 100 varieties~\cite{ampere-gen}), we find both mobile and desktop solutions (e.g., GeForce 30 series~\cite{geforce-30}) and datacenter products (e.g., NVIDIA A100~\cite{a100}). 
Given this variety of devices, we focus on the datacenter class for two reasons: first, this choice reduces variability within the collected metrics associated with product differentiation targeting consumer graphics and display applications (vs. parallel computation in commercial and industrial applications). 
Second, datacenter GPUs are a better indicator of computing innovation, since higher priced products are more likely to embody novel architectural enhancements.

We gathered metrics for all NVIDIA datacenter \acp{GPU} from the Tesla microarchitecture (released in 2006) to the Blackwell Ultra (released in 2025), totaling 102 GPUs.\footnote{Curiously, the term \textit{Tesla} denoted not only the first \ac{GPU} microarchitecture supporting \ac{CUDA} but also the class of general-purpose \acp{GPU} until NVIDIA rebranded it into \textit{datacenter} when releasing Ampere-based devices.}
We focus on on GPU-accelerator cards that require and interact with a host CPU; we do not collect devices such as NVIDIA superchips, which can also function as standalone systems.
\Cref{table:collected_gpus} reports the model names of the GPUs in our dataset, along with the microarchitecture to which they belong and the release year of that microarchitecture.
Specifically, we are interested in metrics describing the \ac{GPU} chip characteristics (e.g., die size and transistor counts),  microarchitecture (e.g., number of CUDA compute units and on-/off-chip memory), and peak theoretical performance at different data precision (e.g., \ac{FP32}). 
For more details about the collected metrics for each \ac{GPU} microarchitecture, see \Cref{table:apx_gpu_metrics} in \Cref{app:collected_metrics}. 

We use the following data sources: online databases such as TechPowerUp~\cite{techpowerup-website} and VideoCardz~\cite{videocardz-website} to create a list of NVIDIA datacenter GPUs; NVIDIA documentation (e.g., whitepapers and datasheets) as the primary source for technical metrics, and TechPowerUp and VideoCardz as secondary sources when the NVIDIA documentation was unavailable; and the aforementioned databases and other websites/datasets (e.g., Epoch AI~\cite{EpochMachineLearningHardware2024}) to collect release prices.
Unfortunately, retrieving accurate and reliable data for the target GPUs was not always possible, particularly for metrics such as initial release price. NVIDIA used to publicly release highly detailed documents describing the features of each GPU microarchitecture or specific GPU models~\cite{p100, a100_whitepaper, h100}. However, the public documentation for the most recent microarchitectures (e.g., Blackwell and Blackwell Ultra) is no longer as exhaustive as it once was, and it tends to focus on specific aspects (e.g., AI-oriented performance) rather than the full range of GPU features. 
Consequently, we infer some values during our data collection process to fill such gaps.\footnote{
For instance, if NVIDIA documentation did not report the Tensor Core \ac{MAC} operations per clock cycle, we computed this value using the number of compute units and running frequency.
Please note that NVIDIA documentation relies on GPU boost clock when reporting the theoretical maximum TFLOPS of CUDA/Tensor Cores at different data precisions. Moreover, Hopper GPUs also have a separate boost clock for Tensor Cores supporting all data precisions but FP64 (which relies on the GPU boost clock). 
The limitation of this approach will be discussed in \secref{sec:limitations}.
Similarly, we extrapolated the features of B100, B200, and B300 GPUs from their respective HGX platforms (the NVIDIA HGX is a platform that integrates and interconnects multiple GPUs to accelerate different workloads~\cite{hgx}).
Another noteworthy detail is that datacenter \acp{GPU} based on desktop/workstation-oriented microarchitectures (e.g., Turing and Ada Lovelace) or boards featuring specific chips (e.g., the GA102 chip~\cite{a102} inside the A40 \ac{GPU}) also include additional graphics components at the cost of other computing features, as also mentioned in \Cref{sec:background_gpu}.
For example, the Ada Lovelace L40 \ac{GPU} features NVIDIA third-generation \ac{RT} Cores and substantially fewer \ac{FP64} \ac{CUDA} Cores. 
Consequently, we included only details relevant to general-purpose or specialized computing, not to graphics.}
Finally, we considered theoretical maximum values for computing and memory performance provided by NVIDIA.
Reaching this level of performance may not always be feasible in real-world computing scenarios, but these theoretical values indicate the peak theoretical potential of the target \acp{GPU} and serve as a consistent performance indicator comparable across products and over time. Furthermore, this approach enables us to decouple and evaluate each GPU component separately, which would be unfeasible with real benchmarks.

In summary, we have built a comprehensive dataset of technical and economic metrics for datacenter GPUs. Although the analysis in this paper leverages a subset of such metrics, we believe our data collection can pave the way for further analyses in the future.

\subsection{Data Selection}\label{subsec:data_selection}
\subsubsection{Metric Selection}\label{subsubsec:metric_selection}
Our analysis identifies relevant trends in GPU progress by looking at the main characteristics defining GPU performance. 
To this end, we focus on theoretical \acf{FP} computing performance, off-chip memory size and bandwidth, release price, and power consumption. 
We aim to analyze trends in GPU improvements from a theoretical peak perspective, without referencing specific benchmarks or AI models, which vary significantly and are unavailable for GPUs released over the entire target period (2007 to 2025). 
This approach allows us to consistently compare progress in different GPU capabilities over decades. We do not distinguish between single- and multi-die GPUs in our analysis, treating the latter as a single logical GPU.
We address the limitations in our analysis in \secref{sec:limitations}.

For computing performance metric, we rely on the theoretical maximum \ac{TFLOPS} that \acp{GPU} can process when employing \ac{FP16}, \ac{FP32}, and \ac{FP64}. 
\ac{FP16} data precision was available only through software emulation before NVIDIA introduced hardware support in the Pascal microarchitecture (2016).
Hence, we use FP32 performance values to represent FP16 performance in previous microarchitectures (from Tesla to Maxwell)~\cite{9623445}.
As described in \Cref{sec:background_gpu}, NVIDIA introduced Tensor Cores in the Volta microarchitecture (2017) for \ac{FP16} tensor computations and later extended the support to \ac{FP32} and \ac{FP64}, as well as other data precisions that we do not analyze in this study (e.g., 8-bit INT/FP precision).\footnote{Technically, Tensor Cores employ the TF32 data format for 32-bit operations. Since we are interested in the theoretical performance that both CUDA and Tensor Cores can deliver for 32-bit operations, we make no distinction between FP32 and TF32 data types and refer to both as FP32 for simplicity.}
In addition, starting from the Ampere microarchitecture, Tensor Cores also benefit from hardware sparsity support when processing inputs featuring a 2:4 sparsity pattern (i.e., for each group of four contiguous values, at least two must be zero~\cite{sparsity}).\footnote{Tensor Cores support sparsity up to 32-bit data precision.}
Conversely, since the number of \ac{FP64} \ac{CUDA} Cores differs dramatically by \ac{GPU} chip or microarchitecture, we sort devices into two categories based on how many \ac{FP64} processing units the model of \ac{GPU} contains, and study each group's evolution separately.
Specifically, we use a threshold of 8 \ac{FP64} \ac{CUDA} Cores per \ac{SM} to discriminate between these two categories.
For instance, the GA100~\cite{a100} and GA102~\cite{a102} GPU chips contain 32 and 2 \ac{FP64} \ac{CUDA} Cores per \ac{SM}, respectively.
Our work does not analyze lower data precisions such as FP8 and FP4. Although they are undoubtedly relevant to fields like AI~\cite{dettmers2022llmint88bitmatrixmultiplication, gupta2015deeplearninglimitednumerical, courbariaux2015trainingdeepneuralnetworks}, Tensor Cores typically execute them, meaning their performance is proportional to that of Tensor Cores for FP16. In other words, when the bitwidth halves (e.g., from 16 to 8 bits), the performance usually doubles (the opposite generally happens when the bitwidth doubles, except for FP64, whose performance decreases by roughly 16$\times$ or more because FP64 operations are not as common as lower-precision operations in the AI field).\footnote{One exception to this pattern is the Blackwell Ultra B300, whose FP4 performance is roughly 3.1$\times$ higher than FP8 performance (14000 TFLOPS vs. 4500 TFLOPS, based on the HGX B300 specifications~\cite{blackwell_technical_overview}).} Consequently, observations for these data precisions could be extrapolated by FP16 trends.

NVIDIA \acp{GPU} mainly feature two types of off-chip memory technologies: \ac{GDDR} or \ac{HBM}. 
Older GPUs and those based on desktop/workstation-oriented microarchitectures mainly use the former technology, while pure datacenter GPUs rely on the latter.
Both types of memory went through steady improvement over the years, offering increasing memory bandwidth and capacity. 
Our work focuses on memory size (or capacity) and theoretical maximum memory bandwidth, reporting them in \ac{GB} and \ac{GBS}, respectively, because these are the main metrics that characterize off-chip memory.
To enhance comparability with published studies of computer and semiconductor price trends undertaken by government statistical agencies and academic researchers, we report original release prices in nominal (current year) US dollars in our analyses (adjustment for price inflation with the reader's price deflator of choice is easily accomplished).
Finally, we use \ac{TDP} as an indicator of the \ac{GPU} power consumption; \ac{TDP} is often about two-thirds of peak power draw, according to one prominent computer architecture textbook~\cite{10.5555/3207796}. 


\subsubsection{Top-Performing GPUs Per Year Selection}\label{subsubsec:top_gpus_selection}
For this subset, we pick the top-performing GPUs per year based on the average of the peak performance in FP16, FP32, and FP64, delivered using either CUDA or Tensor Cores (when available), and for which we were able to find a published release price. 

\subsection{Methodology}\label{subsec:methodology}
After collecting data for the GPUs reported in \Cref{table:collected_gpus} and selecting the metrics of interest, 
our methodology leverages a regression model to analyze the main features of datacenter GPUs. In particular, we fit the data to an exponential growth model:
\begin{equation}\label{eq:exp_regression}
    \widehat{Y} = \alpha e^{(\beta t)}
\end{equation}
where the input variable $t$ indicates the year, $\alpha$ and $\beta$ are parameters in the exponential model, and the dependent variable $\widehat{Y}$ represents the estimated value of a given metrics.
Taking logarithms on both sides of the equation, we have a linear model with the log of the dependent variable as a linear function of time.
The estimated coefficient of time in this linear model can be transformed into point estimates of the \acf{CAGR} — or \ac{AGR}, since they are identical in this constant growth rate model — and \acf{DT} over the entire sample data time period. Approximate robust standard errors for \ac{CAGR} and \ac{DT} were derived from the estimated robust standard errors of the linear model time coefficient using the delta method.\footnote{If $\beta$ is the coefficient of time in this linear model, and $\hat{\sigma_\beta}$ is estimated standard error, \ac{CAGR} is estimated as $\ e^\beta-1$; DT is estimated as $\ln(2)/\beta$. Using the delta method, $\hat{\sigma_\beta}(CAGR) = e^\beta \hat{\sigma_\beta}$ ; $\hat{\sigma}(DT) = (\ln(2)/\beta^2) \hat{\sigma_\beta}$.}
We use Python, Pandas, SciPy, NumPy, and the Statsmodels~\cite{seabold2010statsmodels} statistical modeling and econometrics package to compute the model coefficients and derive the estimated trend model CAGR and DT estimates through the Statsmodels' Ordinary Least Squares (OLS) regression. Moreover, we calculate the standard errors and other sample statistics (e.g., mean and standard deviation) using Scipy's Percent Point Function (PPF) and Pandas' \texttt{describe} function.

Given these selected metrics, we estimated log-linear exponential regression models (Equation~\ref{eq:exp_regression}) using our dataset and derived the \acp{CAGR}. 
Our analysis reported estimates of trend \acp{CAGR} and DTs, along with 90 percent confidence intervals, for all technical improvement metrics, and technical improvement per-memory bandwidth, per-dollar, and per-watt ratios. 
Non-overlapping confidence intervals are a conservative sufficient condition for rejecting the hypothesis that two parameters are actually increasing at the same rate~\cite{Wright03042019}. 
In other words, if two statistics have non-overlapping confidence intervals, they are necessarily significantly different.


\section{GPU Progress Trends}\label{sec:gpu_progress_analysis}

This section analyzes diverse \ac{GPU} metrics to identify trends describing \acp{GPU}' past and current progress. 
We start by examining trends in a subset comprising the top-performing \acp{GPU} per year.
Then, we generalize and expand our study to cover all NVIDIA datacenter \acp{GPU} released from 2007 to 2025, and measure their average improvement as a group.
Finally, we compare and discuss the trends across these two groups of datacenter \acp{GPU}.

\subsection{Top-Performing GPUs Per Year}\label{subsec:top_gpu}
The first objective of our work is to analyze specific trends for the highest performing \acp{GPU} by year.
Specifically, our analysis focuses on trends that highlight the progress of technical metrics, as well as the ratios between computational performance and memory bandwidth, release price, and power consumption.
We start with this focus because we believe it is easier to identify and appreciate changes and improvements throughout the various microarchitectures of NVIDIA \acp{GPU} by concentrating on a small selection of best-in-class devices. Conceptually, these contain the maximum computing power that can be purchased in a single NVIDIA parallel computing device, at any price. For each metric, we compute both the CAGR and DT.

\begin{figure*}[t]
    \centering
    \includegraphics[width=\textwidth]{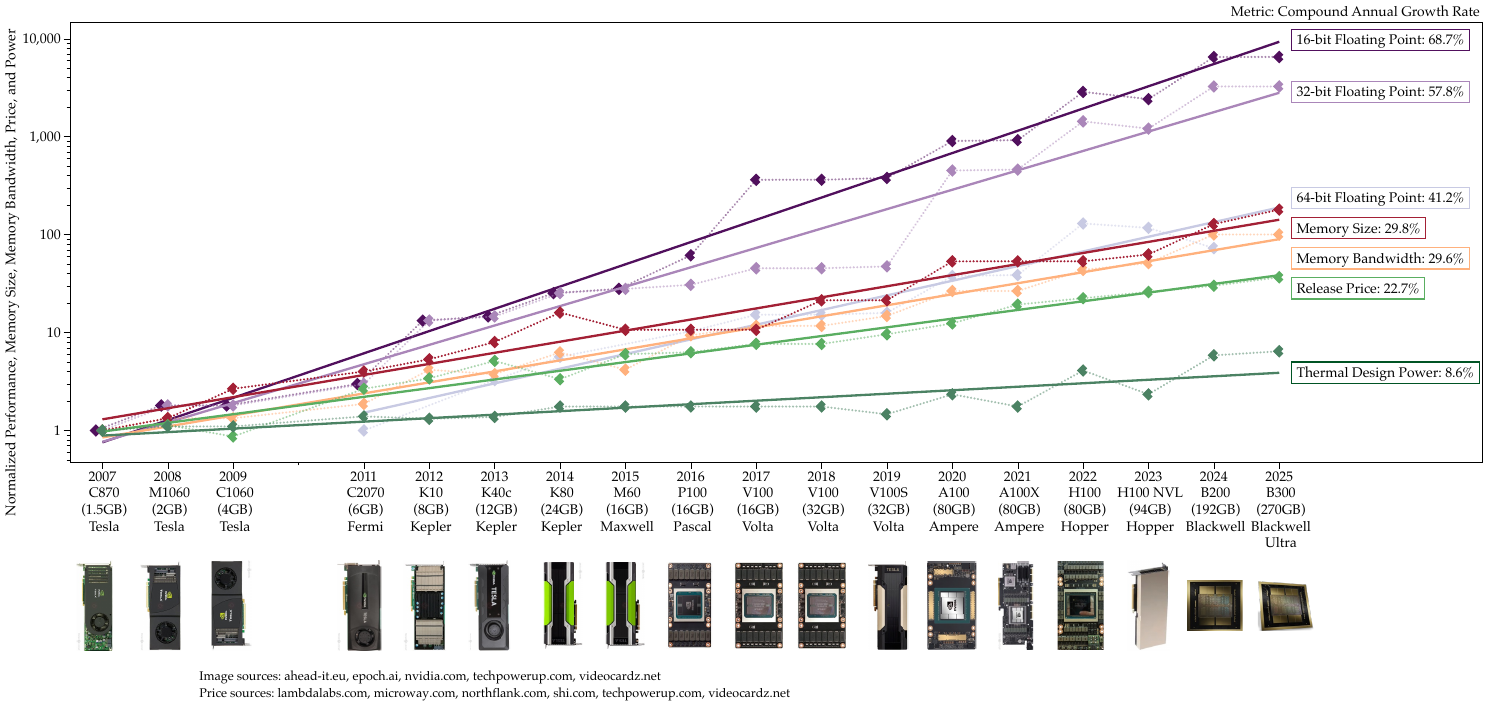}
    \caption{\textbf{Top-performing GPUs per year:} Computing performance, off-chip memory size and bandwidth, release price, and \acf{TDP} scaling normalized to Tesla C870 (2007). 
    For \acf{FP64}, the values are normalized to Fermi C2090 (2011).
    The Figure also reports the estimated \acfp{CAGR} based on exponential trends derived from the selected \acp{GPU} (solid lines) and indicates \ac{FP64} performance for \acp{GPU} with more than 8 \ac{FP64} \ac{CUDA} Cores per \acf{SM}. 
    Finally, the Figure shows the \ac{GPU} images and reports the off-chip memory size to help identify the specific model.}
    \label{fig:top_gpu_scaling}
\end{figure*}

Given the subset of top-performing GPUs per year, we made the following assumptions and considerations to analyze major trends.
Since Tensor Cores outperform \ac{CUDA} Cores in tensor computations and the \ac{TFLOPS} of these two processing unit types do not sum up~\cite{cuda_and_tensor_cores}, we use the former as the performance reference when available.
In this way, we can observe the peak GPU performance and the transition from one type of computing resource to another.
Then, to keep this analysis generalizable to a larger class of computational workloads, we do not take into account hardware sparsity support when available, and just report \ac{FP64} performance for \acp{GPU} with numbers of \ac{FP64} \ac{CUDA} Cores per \ac{SM} exceeding 8. 

\begin{figure*}[t]
\centering
\includegraphics[width=0.92\textwidth]{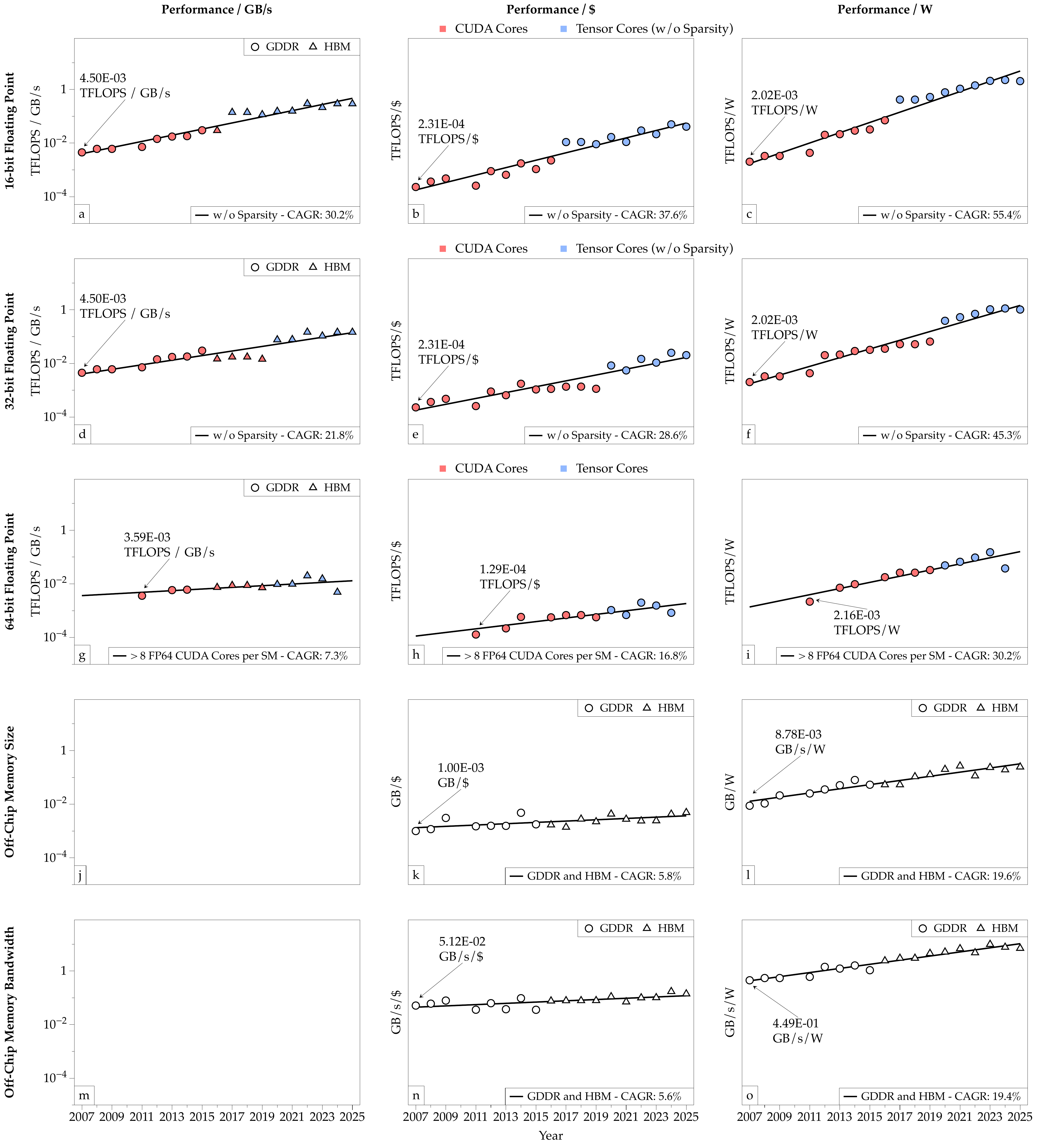} 
    \caption{\textbf{Top-performing GPUs per year:} A series of plots reporting the exponential scaling trends for the technical improvements per memory bandwidth (left column), per dollar (central column), and per watt (right column). Each subplot also contains the \acf{CAGR} of the given trend. Subplots \textit{a} through \textit{i} use colors to indicate whether the performance values are derived from CUDA Cores (red) or Tensor Cores without sparsity (blue). Similarly, subplots \textit{a}, \textit{d}, \textit{g}, \textit{k}, \textit{l}, \textit{n}, and \textit{o} use shapes to indicate whether the off-chip memory size and bandwidth values are derived from GDDR (circle) or HBM (triangle) technology.}
    \label{fig:top_gpu_ratio_charts}
\end{figure*}

\subsubsection{Technical Improvements}\label{subsubsec:top_technical_improvements}
\Cref{fig:top_gpu_scaling} shows trend lines for computational performance (\ac{FP16}, \ac{FP32}, and \ac{FP64}), off-chip memory size and bandwidth, release price, and \ac{TDP} metrics normalized to the NVIDIA Tesla C870 from 2007 (or the NVIDIA Fermi C2070 from 2011 for \ac{FP64}).
Introducing FP16 hardware support first (from the 2016 Pascal microarchitecture) and Tensor Cores later (from the 2017 Volta microarchitecture) boosted the performance on \ac{FP16} by roughly an order of magnitude. 
The 68.7\% \ac{CAGR} for FP16 performance, which implies a 1.32y DT, is only a little less than trend growth rates measured for Intel desktop processor performance (an average measured for all desktop CPUs, using a similar methodology) in the late 1990s, at the peak of Moore's Law/Dennard scaling-driven performance improvement~\cite{flamm2021measuring}. 
The trend for \ac{FP32} performance is similar, especially after Tensor Core support was introduced in 2020 (Ampere microarchitecture). 
We also observe a 41.2\% CAGR for FP64, which is relatively lower than that of FP16 and FP32 and implies an estimated DT of almost 2 years. Besides, after a peak in 2022 (Hopper H100), FP64 performance has been decreasing in subsequent years, reaching a point where top-notch GPUs like the Blackwell Ultra B300 have fewer than 8 FP64 CUDA Cores per SM (which is why we do not show FP64 performance for B300 in \Cref{fig:top_gpu_scaling}).
Moving to other metrics, the introduction of HBM technology in GPUs (starting with the 2016 Pascal microarchitecture) significantly contributed to the growth of off-chip memory, with sharp increases in both size (29.8\% CAGR and 2.66y DT) and bandwidth (29.6\% CAGR and 2.68y DT).
Price at launch, the cost of purchasing \acp{GPU} has been consistently increasing over time, at a 22.7\% \ac{CAGR} from 2007 to 2025 (3.39y DT).
Finally, the \ac{TDP} of the selected NVIDIA \acp{GPU} grew by a much slower 8.6\% every year (8.45y DT), as the 170.9W (Tesla C870) evolved into the 1100W (Blackwell Ultra B300).

\subsubsection{Technical Improvement Ratios}
After analyzing the technical improvements in the top-performing GPUs per year, we now focus on their per-memory bandwidth, per-dollar, and per-watt ratios. 
\Cref{fig:top_gpu_ratio_charts} reports the exponential scaling trends for these metrics and also indicates when the values represent CUDA/Tensor Core performance (point color in subplots \textit{a} through \textit{i}) or GDDR/HBM size/bandwidth (point shape in subplots \textit{a}, \textit{d}, \textit{g}, \textit{k}, \textit{l}, \textit{n}, and \textit{o}).

Starting with the technical improvements per memory bandwidth (left column of plots in \Cref{fig:top_gpu_ratio_charts}), we consider only the ratio between computational performance (i.e., FP16, FP32, and FP64) and off-chip memory bandwidth, as the memory size per-bandwidth ratio essentially translates into the memory transfer time. 
The FP16 per-bandwidth trend data points show the benefits Tensor Cores have provided since 2017, compensating for bandwidth improvements and resulting in a 30.2\% CAGR (2.62y DT). 
Such a pattern is even more evident in FP32, where we can observe how HBM had been affecting trend scaling between 2016 and 2019, until the introduction of Tensor Core support for FP32 (or, rather, TF32), producing a 21.8\% CAGR, which implies a 3.52y DT. 
Finally, the trend curve for FP64 is relatively flat (7.3\% CAGR and 9.79y DT), mainly because its growth has been slower than that of FP16 and FP32.
These trends generally indicate that computational performance growth tends to outpace off-chip memory bandwidth, potentially implying that the latter may become a bottleneck for computations that depend heavily on memory transfers.
Consequently, this memory or bandwidth wall~\cite{10.1145/1555815.1555801} would prevent GPUs from reaching their peak computational performance.

The central column of plots in \Cref{fig:top_gpu_ratio_charts} illustrates the technical improvements per dollar. Starting with computational performance, we observe higher CAGRs than in the per-memory bandwidth trends. Moreover, the patterns and gaps between growth rates remain pretty similar to what we analyzed before. On the other hand, the scaling trends based on memory characteristics (i.e., size and bandwidth) per dollar are moderately flat and yield similar CAGRs (5.8\% and 5.6\%, respectively).

Finally, \Cref{fig:top_gpu_ratio_charts} (right column of plots) depicts the exponential scaling trends of the technical improvements per watt.
In general, we notice again patterns similar to those of the previous ratios. However, this time, the growth rates of these trends are notably higher in both computational performance and memory characteristics, due to the slower growth of TDP. Specifically, the CAGRs (DTs) range from 55.4\% (1.57y) to 19.4\% (3.92y) for FP16 and memory bandwidth, respectively. The gaps in growth rates across the different ratios remain roughly consistent.

\subsection{All Datacenter GPUs}

We now expand and generalize these analyses to the entire population of GPUs in our curated dataset and examine trends across all selected metrics, also considering different hardware feature configurations when available.
Specifically, we focus on technical improvements for computing performance, off-chip memory size and bandwidth, release price, and TDP, as well as the per-memory bandwidth, per-dollar, and per-watt ratios.
Even in this case, we estimated log-linear exponential regression models (Equation~\ref{eq:exp_regression}) and calculated \acp{CAGR} and DTs.
For the performance metric, each data point represents the maximum value in terms of \ac{TFLOPS} that a \ac{GPU} can deliver at a given data precision. 
As we did in \secref{subsec:top_gpu}, we replace \ac{CUDA} Cores with Tensor Cores in the performance metrics as soon as the latter units become available to focus on peak theoretical performance at a given precision.
On the other hand, the sparsity support within Tensor Cores (introduced in the Ampere microarchitecture) is a data-dependent feature for data precisions up to 32 bits.
For this reason, we report separate exponential trend lines with and without the theoretical 2$\times$ performance improvements granted by sparsity. 
Instead, we fit performance for 64-bit data precision using two trend lines based on the number of FP64 CUDA Cores per SM to account for the significant performance variability across datacenter GPUs with desktop/workstation-oriented microarchitectures. Even in this analysis, we set the threshold to 8 FP64 CUDA Cores per SM.
Similarly, to assess the separate contributions of GDDR and HBM to off-chip memory size and bandwidth scaling, we show decoupled trend lines for GDDR, HBM, or both.
Finally, we study the exponential scaling of release prices and TDP.

\subsubsection{Technical Improvements}

\begin{figure*}[t]
    \centering
    \includegraphics[width=0.83\textwidth]{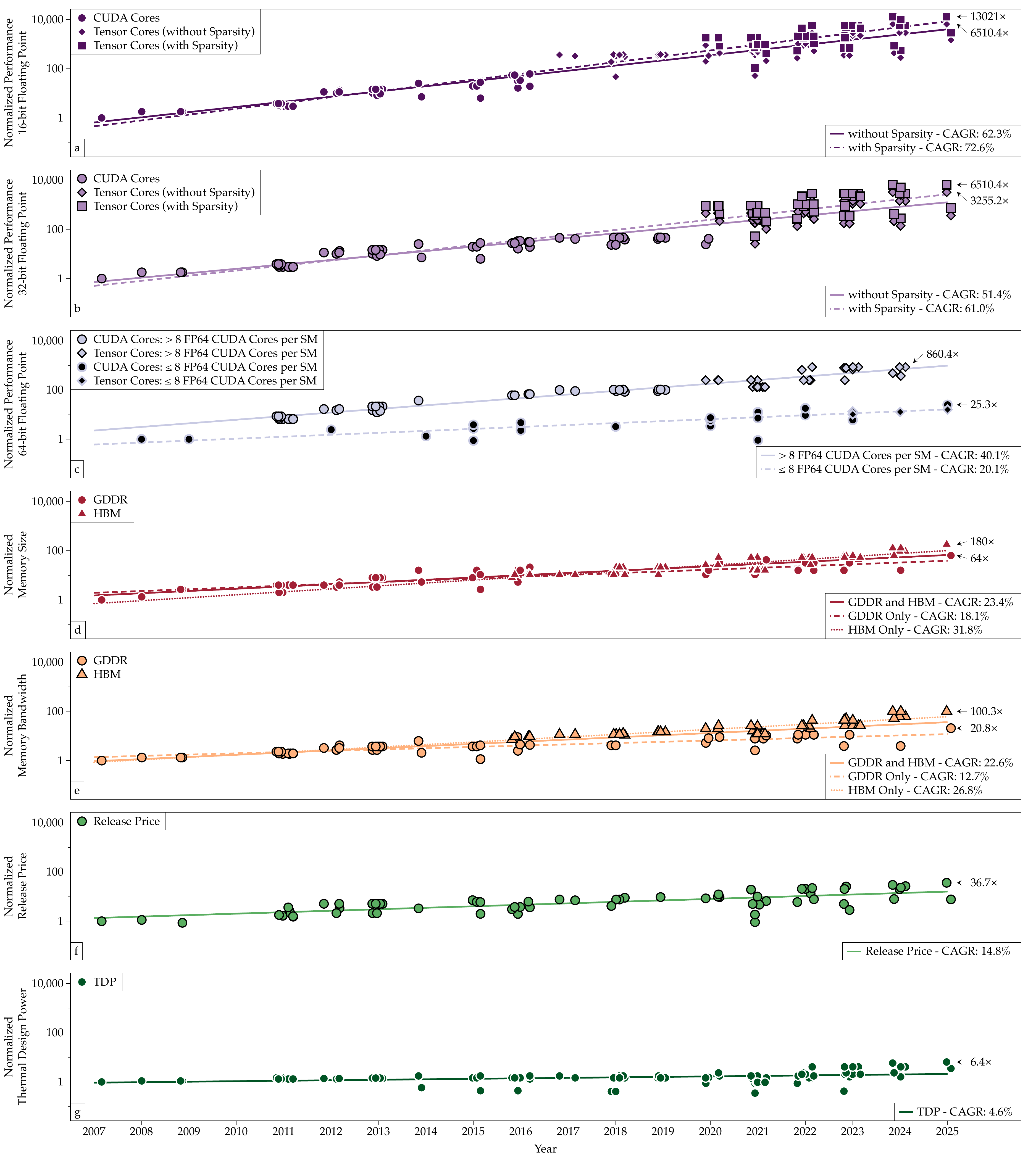}
    \caption{\textbf{All datacenter GPUs:} A series of plots reporting the exponential scaling trends for computing performance (subplots \textit{a} through \textit{c}), off-chip memory size and bandwidth (\textit{d} and \textit{e}), release price (\textit{f}), and \acf{TDP} (\textit{g}). The values are normalized to Tesla C870 (2007) for all metrics except \acf{FP64}, which are normalized to Tesla M1060 (2008) or Fermi C2050 (2011), depending on the number of FP64 CUDA Cores per \acf{SM}. Each subplot contains one or more trends for each metric configuration, along with their \acfp{CAGR}. Subplots \textit{a} through \textit{c} use shapes to indicate whether the performance values are derived from CUDA Cores (circle) or Tensor Cores, without (diamond) or with (square) sparsity. In addition, subplot c colors either the center or the border of each point according to the amount of FP64 units per SM. Similarly, subplots \textit{d} and \textit{e} use shapes to indicate whether the off-chip memory size and bandwidth values are derived from GDDR (circle) or HBM (triangle) technology. Finally, we apply horizontal jitter to prevent overplotting.}
    \label{fig:all_gpu_scaling}
\end{figure*}

\Cref{fig:all_gpu_scaling} illustrates how the technical improvements affect the estimated \acp{CAGR} of \ac{FP16}, \ac{FP32}, \ac{FP64}, off-chip memory size and bandwidth, release price, and \ac{TDP} for all NVIDIA datacenter \acp{GPU}.
\Cref{fig:all_gpu_scaling} highlights the impact of sparsity (for \ac{FP16} and \ac{FP32}), the amount of \ac{FP64} CUDA Cores, and different types of off-chip memory technology on the growth rates.
Starting with performance metrics, \ac{FP16} and \ac{FP32} show impressive \acp{CAGR} (62.3\% and 51.4\% in subplots \textit{a} and \textit{b}), which imply DTs of 1.43 years and 1.67 years, respectively.
Sparsity support (dashed lines), if it can be utilized, further increases the \ac{CAGR} by roughly 10 \acfp{PPT}.
On the other hand, FP64 performance growth varies significantly with the number of CUDA Cores per \ac{SM} (from 20.1\% to 40.1\% in \Cref{fig:all_gpu_scaling}.c), reflecting differences in the allocation of computing resources for this data precision across GPU microarchitectures, from pure datacenter to desktop/workstation-oriented ones.
Moving to off-chip memory, we see the separate and joint contributions of GDDR and HBM technologies to the growth in memory size and bandwidth (subplots \textit{d} and \textit{e}). Since its introduction in 2016 (Pascal P100), HBM (dotted lines) impacted both memory size (31.8\% CAGR) and bandwidth (26.8\% CAGR), outpacing GDDR (dashed lines) by roughly 14 \acp{PPT}.
HBM immediately boosted memory bandwidth, as the top HBM-based bandwidth from Pascal microarchitecture (732.2GB/s in the 2016 Pascal P100 DGXS) exceeds the top GDDR-based bandwidth from the previous microarchitecture (332.8GB/s in the 2016 Maxwell M10) by more than 2$\times$. Conversely, the initial size/capacity of HBM was smaller than that of GDDR (16GB in the 2016 Pascal P100 DGXS vs. 32GB in the 2016 Maxwell M10), so it took longer for HBM to surpass GDDR in terms of memory size. Consequently, when we analyze the joint contribution of GDDR and HBM (solid lines), we can observe that the memory size CAGR (23.4\%) is closer to that of GDDR, whereas the memory bandwidth CAGR (22.6\%) is closer to that of HBM.

Release prices indicate a 14.8\% CAGR and a consequent 5.03y DT (\Cref{fig:all_gpu_scaling}.f).\footnote{The limitations of this analysis will be discussed in \secref{sec:limitations}.}
Finally, as the maximum power that can be delivered to a PCIe-based card is $\sim$300W (power demand over that requires more specialized/custom power supply, cabling, etc.), many GPUs in our dataset generally share similar TDP values ranging from 200W to 300W (60 GPUs out of 102), yielding a particularly slow growth rate for this metric (4.6\% in \Cref{fig:all_gpu_scaling}.g).

\begin{figure*}[ht]
\centering
\includegraphics[width=0.92\textwidth]{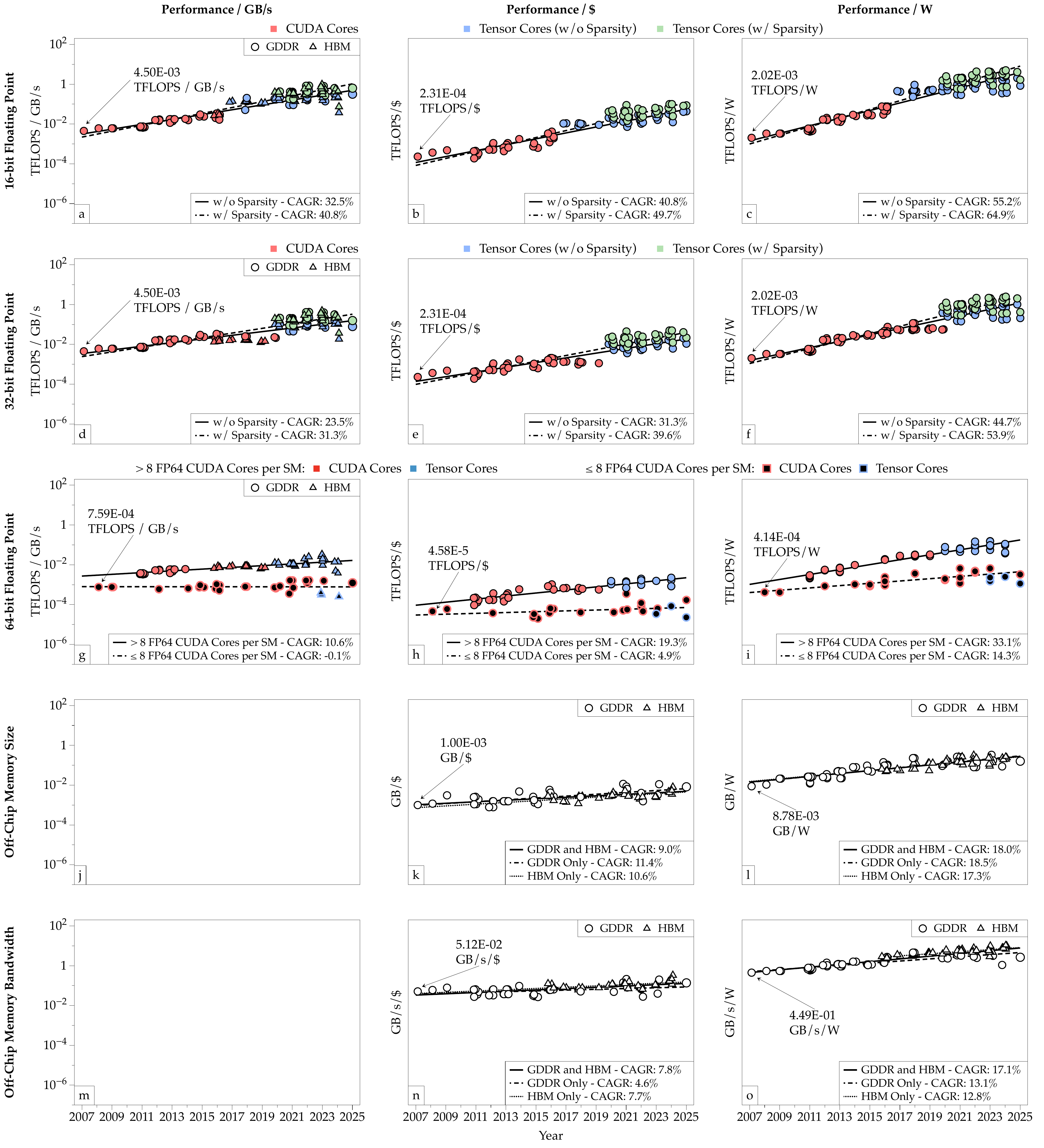} 
    \caption{\textbf{All datacenter GPUs:} 
    A series of plots reporting the exponential scaling trends for the technical improvements per memory bandwidth (left column), per dollar (central column), and per watt (right column). Each subplot contains one or more trends for each metric configuration, along with their \acfp{CAGR}. Subplots \textit{a} through \textit{i} use colors to indicate whether the performance values are derived from CUDA Cores (red) or Tensor Cores, without (blue) or with (green) sparsity. In particular, subplots \textit{g} through \textit{i} color either the center or the border of each point according to the amount of \acf{FP64} units per \acf{SM}. Similarly, subplots \textit{a}, \textit{d}, \textit{g}, \textit{k}, \textit{l}, \textit{n}, and \textit{o} use shapes to indicate whether the off-chip memory size and bandwidth values are derived from GDDR (circle) or HBM (triangle) technology. Finally, we apply horizontal jitter to prevent overplotting.}
    \label{fig:all_gpu_ratio_charts}
\end{figure*}

\subsubsection{Technical Improvement Ratios}
\Cref{fig:all_gpu_ratio_charts} depicts the exponential scaling trends \textit{for all datacenter GPUs} under different configurations, which include CUDA/Tensor Core performance with and without sparsity (point color in subplots from \textit{a} through \textit{i}), amount of FP64 CUDA Cores per SM (inner and outer point color in subplots from \textit{g} through \textit{i}), and GDDR/HBM size/bandwidth (point shape in subplots \textit{a}, \textit{d}, \textit{g}, \textit{k}, \textit{l}, \textit{n}, and \textit{o}).

The left column of plots in \Cref{fig:all_gpu_ratio_charts} shows the technical improvements per memory bandwidth. This specific analysis considers different configurations that affect computational performance (e.g., using CUDA Cores vs. Tensor Cores, with or without sparsity), as these represent choices end users can make for their workloads. 
As HBM has been replacing GDDR as the dominant off-chip memory technology for datacenter-oriented GPU microarchitectures over the years, our analysis of technical improvements per memory bandwidth does not decouple the contributions of GDDR and HBM to highlight this transition.
Nonetheless, the Figure still indicates what type of off-chip memory each point derives from. The trends for the FP16 per-memory bandwidth ratio yield CAGRs of 32.5\% (without sparsity) and 40.8\% (with sparsity), driven mainly by the performance boost that Tensor Cores and sparsity support provide. The Tensor Core contribution is particularly evident in the FP32 per-memory bandwidth trends (CAGRs of 23.5\% without sparsity and 31.3\% with sparsity), as the scaling of this metric with CUDA Cores (red points) plateaued before the introduction of Tensor Cores. Finally, the two trend lines for the FP64 per-memory bandwidth ratios exhibit intriguing results. While GPUs with a high number of FP64 CUDA Cores and consequently good computational capabilities yield an 10.6\% CAGR and 6.90y DT (solid line in \Cref{fig:all_gpu_ratio_charts}g), the other subset of GPUs shows a decreasing trend (dashed line in \Cref{fig:all_gpu_ratio_charts}g), suggesting that memory bandwidth outpaces the performance delivered by this configuration.

The central column of plots in \Cref{fig:all_gpu_ratio_charts} reports the technical improvements per dollar. Since the growth of release prices is not as significant as that of off-chip memory bandwidth (see \Cref{fig:all_gpu_scaling}), these trends generally yield better CAGRs than those observed in the left column of plots in \Cref{fig:all_gpu_ratio_charts}. In particular, the trends in computational performance for FP16 and FP32 indicate CAGRs of 40.8\% (49.7\% with sparsity) and 31.3\% (39.6\%), respectively, roughly 7 to 8 \acp{PPT} higher than those observed in the per-memory bandwidth ratios. In addition, both subsets of GPUs, based on the number of FP64 CUDA Cores per SM, show positive growth trends, even though the growth for low FP64 capabilities is relatively slow (4.9\% CAGR and 14.36y DT). Moving to off-chip memory, it is interesting to note that the three trend lines for each metric (memory size and bandwidth) almost overlap. Indeed, these per-dollar ratios significantly reduced the gap and contribution of HBM technology we highlighted in \Cref{fig:all_gpu_scaling}, mainly due to the higher release price of HBM-equipped GPUs. Consequently, the CAGRs range from 9.0\% (GDDR and HBM) to 11.4\% (GDDR only) for off-chip memory size and from 4.6\% (GDDR only) to 7.8\% (GDDR and HBM) for bandwidth.

Finally, we discuss the technical improvements per watt (right column of plots in \Cref{fig:all_gpu_ratio_charts}). The TDP growth varies less with computational performance across different precisions than the release price and off-chip memory. Therefore, the resulting trends for FP16, FP32, and FP64 are just 5.8 ($\le$8 of FP64 CUDA Cores per SM) to 7.7 (FP16 with sparsity) \acp{PPT} away from those in \Cref{fig:all_gpu_scaling}, implying DTs ranging from 1.39 to 5.19 years. On the other hand, we again notice a narrow gap between the configurations we selected for the metrics based on off-chip memory size and bandwidth. Generally, HBM-based GPUs have higher TDPs than GDDR-based ones (compensating for higher memory size and bandwidth), which explains why the separate CAGRs for GDDR and HBM are so close. Instead, the joint contribution of GDDR and HBM yields CAGRs that are both between and higher than the separate ones for the off-chip memory size and bandwidth metrics, respectively.
This result further demonstrates how HBM has contributed to the progress in both off-chip memory size and bandwidth since its introduction in NVIDIA GPUs, as previously mentioned in \secref{subsubsec:top_technical_improvements}, paving the way for a more effective execution of memory-intensive workloads.

\subsection{Trend Comparison}
We now compare the trends identified in both the top-GPU-per-year subset and the full dataset of datacenter GPUs. \Cref{fig:cagr_histogram} summarizes the CAGRs with 90 percent confidence intervals and DTs for technical, per-memory bandwidth, per-dollar, and per-watt improvements for top-performing GPUs (left side of the chart) and all datacenter GPUs (right side). In addition, \Cref{table:apx_data_stats} in \Cref{app:data_stats} presents the main statistics for the data used in our trend analysis, including sample size, mean, standard deviation, and R\textsuperscript{2}.

Starting with the technical improvements, the subset of top-performing GPUs reports higher CAGRs than the whole dataset of datacenter GPUs across all considered configurations, except for one (i.e., $>$ 8 FP64 CUDA Cores per SM), even though the CAGRs are pretty close. Specifically, computational performance metrics are apart by 6.4, 6.4, and 1.1 \acp{PPT} for FP16, FP32, and FP64, respectively. 
On the off-chip memory side, the gaps between the two sets of GPUs range from 6.4 \acp{PPT} (size) to 7 \acp{PPT} (bandwidth).
Finally, the subset of top-performing GPUs increases the CAGRs for release price and TDP by 1.53$\times$ and 1.87$\times$ (or decreases the DTs by 0.67$\times$ and 0.55$\times$), respectively. These results align with our expectations, showing that best-in-class GPUs better reflect the steeper growth in such metrics, even when this outcome implies higher prices or power consumption than the other GPUs from the same period of time.


In terms of per-memory bandwidth, per-dollar, and per-watt ratios, however, we observe a quite different scenario: most metrics yield lower CAGRs when restricting the analysis to the subset of top-performing GPUs. 
In general, this outcome occurs because top GPUs usually feature off-chip memory with higher bandwidth and are the most expensive and power-hungry, as the mean statistic indicates in \Cref{table:apx_data_stats} in \Cref{app:data_stats}, offsetting the performance advantages reflected in the base metric.
For instance, when we analyze the technical improvement per memory bandwidth, the CAGRs range from 7.3\% to 30.2\% for the top-performing GPUs and from -0.1\% to 40.8\% to for the datacenter GPUs, creating a gap ranging from 1.7 to 3.3 \acp{PPT} in favor of the whole GPU dataset across the shared configurations (i.e., FP32 without sparsity and FP64 with $>$ 8 FP64 CUDA Cores per SM).
Besides, we observe a gap spanning from 2.2 to 3.2 \acp{PPT} for CAGRs (or 0.14 to 4.25 years for DTs) when focusing on technical improvement per dollar, another case where the entire GPU dataset exhibits better trends. 
Finally, the analysis of the technical improvement per watt showcases conflicting results. Specifically, the subset of top-performing GPUs shows better trends across all shared configurations except FP64 with $>$ 8 FP64 CUDA Cores per SM, even though the growth rates are quite close for FP16 and FP32 without sparsity.


\section{Top-Performing GPUs Across Vendors}\label{sec:vendor_comparison}
After analyzing progress and trends in NVIDIA datacenter GPUs, we now evaluate how NVIDIA's best-in-class devices compare with those of other vendors across computational performance, off-chip memory size and bandwidth, and power consumption metrics. To this end, we selected the top-performing GPUs per year by NVIDIA, AMD, and Intel from 2017 onwards. We chose to start this comparison from 2017 because it is the year NVIDIA introduced its Tensor Cores and AMD released its first Instinct GPUs (i.e., AMD’s datacenter GPU class~\cite{amd_instinct}). Furthermore, we employed the same approach introduced in \Cref{subsubsec:top_gpus_selection} to select the top-performing GPU per year, with the only difference that we ignored the release price this time, as the following comparison is purely based on architectural features. For this reason, the subset of NVIDIA GPUs in this comparison differs slightly from that employed in \Cref{subsec:top_gpu} (i.e., we replaced the Hopper H100 NVL 94GB with the Hopper H100 SXM 94GB in 2023). Finally, we assembled datasets from AMD and Intel through the following sources: AMD~\cite{amd_datacenter_gpu_list} and Intel~\cite{intel_datacenter_gpu_list} websites and TechPowerUp~\cite{techpowerup-website} to create the list of their datacenter GPUs; the vendor documentation as the primary source for technical metrics, and TechPowerUp as a secondary source when the official documentation was unavailable, just like we did for NVIDIA GPUs. \Cref{table:gpus_per_vendor} reports the list of selected datacenter GPUs per vendor.

\begin{table*}[t]
    \caption{Selected datacenter \acp{GPU} per vendor}
    \label{table:gpus_per_vendor}
    \centering
    \resizebox{\textwidth}{!}{
    \begin{tabular}{cccc}
    \toprule
        \multirow{2}{*}{\textbf{Release Year}} & \multicolumn{3}{c}{\textbf{Microarchitecture | GPU Model | Form Factor | Off-chip Memory Size}}\\
          & \textbf{NVIDIA} &  \textbf{AMD} & \textbf{Intel}\\
        \midrule
        2017 & Volta V100 SXM 16GB & Vega 10 Instinct MI25 PCIe 16GB & - \\
        2018 & Volta V100 SXM 32GB & Vega 20 Instinct MI60 PCIe 32GB & - \\
        2019 & Volta V100S PCIe 32GB & - & - \\
        2020 & Ampere A100 SXM 80GB & CDNA 1 Instinct MI100 PCIe 32GB & - \\
        2021 & Ampere A100X PCIe 80GB & CDNA 2 Instinct MI250X OAM 128GB & - \\
        2022 & Hopper H100 SXM 80GB & CDNA 2 Instinct MI210 PCIe 64GB & Xe-HPG Flex 170 PCIe 16GB \\
        2023 & Hopper H100 SXM 94GB & CDNA 3 Instinct MI300X OAM 192GB & Xe-HPC Max 1550 OAM 128GB \\
        2024 & Blackwell B200 SXM 192GB & CDNA 3 Instinct MI325X OAM 256GB & Xe-HPG Flex 170V PCIe 16GB \\
        2025 & Blackwell Ultra B300 SXM 270GB & CDNA 4 Instinct MI355X OAM 288GB & - \\
        \bottomrule
    \end{tabular}
    }
\end{table*}

\subsection{Architectural Overview}
Before diving into the comparison, we provide a brief, high-level overview of the architectures of AMD and Intel datacenter GPUs to help readers familiarize themselves with these new concepts (an exhaustive architectural description of AMD/Intel devices is out of scope for this paper). As the architectural landscape is embracing the chiplet paradigm, modern NVIDIA, AMD, and Intel datacenter GPUs feature one or more homogeneous/heterogeneous dies (e.g., for computing, memory, and communication) packed together into a single logical processor. At the microarchitecture level, these GPUs are hierarchically organized into parallel collections of computing clusters, which NVIDIA calls \acp{SM}, AMD calls \acp{CU}, and Intel calls X\textsuperscript{e} Cores. Each cluster contains several heterogeneous processing cores or \acp{ALU} that run in \ac{SIMT} mode. Just as NVIDIA GPUs first featured CUDA Cores for scalar operations and later introduced Tensor Cores for matrix/tensor operations, other vendors adopted a similar approach: AMD GPUs contain Stream Processors and Matrix Cores, while Intel GPUs comprise Vector Engines and Matrix Engines. At first approximation, these cores, processors, and engines are comparable from a computational perspective, with the main exception being Intel’s Vector Engines, since a single Vector Engine is a 512-bit wide \ac{ALU} (in the X\textsuperscript{e} architecture~\cite{xe_architecture}) and can be compared to a group of CUDA Cores or Stream Processors. Furthermore, as we observed with NVIDIA GPUs, while scalar/vector units mainly process FP16/BF16, FP32, FP64, and INT32, tensor/matrix units target a broad range of data precisions (with a particular focus on low-bit ones due to their relevance to AI workloads) through various data formats, such as the \ac{OCP} Microscaling formats~\cite{rouhani2023microscaling, mx_specification}, and also exploit native support for sparse data to further boost performance.\footnote{As we did for the analysis of NVIDIA datacenter GPUs, our comparison examines peak theoretical performance at different precisions without differentiating among specific data formats for simplicity.}
Finally, all the AMD GPUs in this comparison employ HBM for their off-chip memories, whereas Intel GPUs still alternate between GDDR and HBM.

\subsection{Vendor Comparison}
\begin{figure*}[t]
\centering
\includegraphics[width=0.92\textwidth]{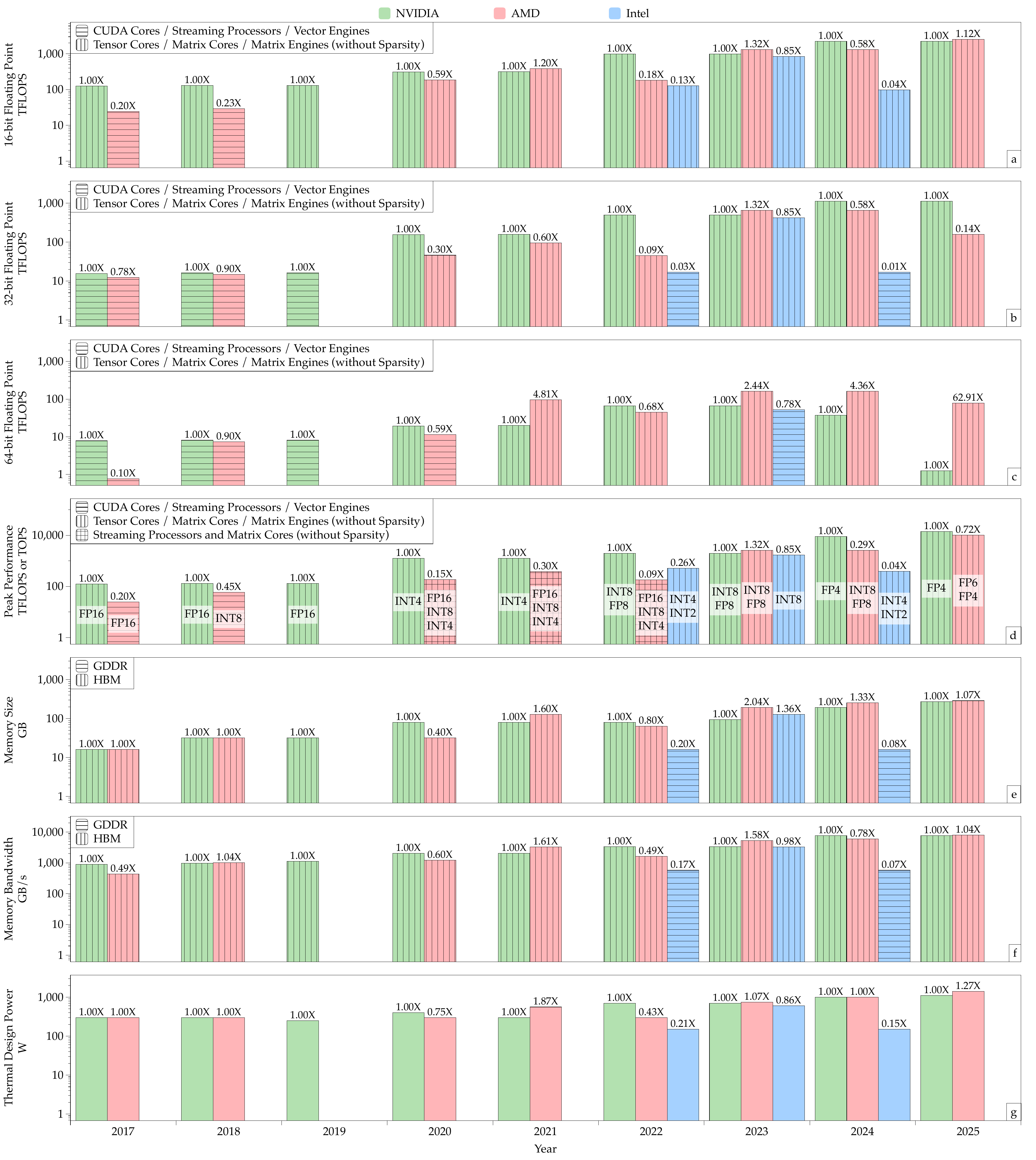} 
    \caption{\textbf{Vendor comparison:} 
    A series of bar plots comparing the top-performing datacenter GPUs from NVIDIA (green bars), AMD (red), and Intel (blue) over the years across \acf{FP16}, \acf{FP32}, and \acf{FP64}, as well as peak computing performance at any precision (subplots \textit{a} through \textit{d}), off-chip memory size and bandwidth (\textit{e} and \textit{f}), and \acf{TDP} (\textit{g}).
    Each bar represents the raw value of a given metric and reports at its top the normalized value relative to the NVIDIA GPU released that year.
    The patterns indicate whether a given computing performance metric is derived from CUDA/Vector Cores or Tensor/Matrix Cores; the same applies to off-chip memories based on GDDR or HBM. The peak performance bars are labeled with the data precision(s) used to derive their values. 
    }
    \label{fig:vendor_comparison_bar_charts}
\end{figure*}

\Cref{fig:vendor_comparison_bar_charts} compares the top-performing NVIDIA (green bars), AMD (red), and Intel (blue) datacenter GPUs per year across computing performance (subplots \textit{a} through \textit{d}), off-chip memory size and bandwidth (\textit{e} and \textit{f}), and TDP (\textit{g}). While each bar indicates the raw value of a given metric, we also report, at the top, the metric value normalized to the NVIDIA GPU released that year, to highlight a factor of improvement ($>$ 1$\times$) or decrease ($<$ 1$\times$) relative to the NVIDIA baseline.

Starting with FP16 performance (\Cref{fig:vendor_comparison_bar_charts}a), after the initial advantage provided by NVIDIA’s Tensor Cores, which gave an improvement of 5.1$\times$ over AMD’s Streaming Processors in 2017, AMD closed the performance gap thanks to the introduction of its Matrix Cores, surpassing NVIDIA by 1.12$\times$ in 2025. 
On Intel’s side, the Max 1550 GPU from 2023 achieved FP16 performance similar to the NVIDIA H100 from the same year, but the gap widened with the following GPU, namely the Flex 170V from 2024, which is more of a general-purpose high-performance device (for instance, it features hardware-accelerated ray tracing support). For this reason, we will use the Intel Max 1550 GPU as the primary Intel device for comparisons going forward. \Cref{fig:vendor_comparison_bar_charts}b shows that NVIDIA and AMD devices had quite close PF32 performance before the introduction of their respective tensor/matrix operation cores.
After that, NVIDIA’s FP32 performance values generally increased, while AMD’s fluctuated, resulting in a 7.15$\times$ gap in favor of NVIDIA in 2025. For Intel, we observe a pattern similar to that of FP16. 
Next, AMD offers significantly higher FP64 performance than the other vendors, achieving 62.91$\times$ and 3.12$\times$ improvements over NVIDIA (2025) and Intel (2023), respectively (\Cref{fig:vendor_comparison_bar_charts}c). 
Finally, when we compare peak performance across data precisions (reported on each bar of \Cref{fig:vendor_comparison_bar_charts}d), we see that the vendors expanded their hardware support to lower-bit precisions and increased the delivered performance. NVIDIA devices generally outperform the other vendors in this metric, even though AMD GPUs are closing the gap.

Moving to the other metrics, we generally observe values that are reasonably similar for off-chip memory size (\Cref{fig:vendor_comparison_bar_charts}e) and bandwidth (\Cref{fig:vendor_comparison_bar_charts}f), especially between NVIDIA and AMD, as they use the same type of off-chip memory, namely HBM. Indeed, the only exceptions to this pattern are the GDDR-based Intel GPUs from 2022 and 2024. Similarly, power consumption values based on the TDP are fairly consistent across years (\Cref{fig:vendor_comparison_bar_charts}g), with AMD devices slightly more power-hungry than NVIDIA’s (e.g., the AMD MI355X has 1.27$\times$ higher TDP than the NVIDIA B300).

NVIDIA gained a significant initial performance advantage over the other vendors thanks to the early introduction of Tensor Cores in 2017 for FP16 computations (and later for other data precisions). AMD and Intel introduced cores/engines for matrix operations in 2020 and 2022, respectively, and immediately targeted a broader range of data precisions than NVIDIA in 2017, mainly following the trends and demand in the AI field. Over the years, the other vendors have reduced the initial FP16 and peak performance gaps with NVIDIA, while we observe conflicting results for FP32 and FP64, highlighting how each vendor invests differently in these precisions, which are not a priority in the current AI-oriented computing domain. On the other hand, adopting the same off-chip memory technology (i.e., HBM) implies reasonably similar sizes and bandwidths, and even the power consumption of the top-performing GPUs from each vendor aligns most of the time. In conclusion, even though the gap is narrowing or has turned around in some cases, the resulting margin is not sufficient to significantly undermine NVIDIA’s dominant market position or justify investments in alternative hardware and software technologies. Indeed, even though an analysis of the software stacks by each vendor is out of the scope of this paper, a remarkable part of NVIDIA’s success is due to CUDA and its libraries, which remain ahead of AMD’s ROCm~\cite{rocm} and Intel’s oneAPI~\cite{oneapi}, in terms of efficiency, diffusion, and support by third-party frameworks~\cite{cuda_vs_rocm_1,cuda_vs_rocm_2}.

\section{Export Control Regulations}\label{sec:export}

As discussed in \secref{sec:gpu_progress_analysis}, the computing and memory performance of NVIDIA \acp{GPU} has been growing at impressive rates, addressing the ever-increasing demand for performance. 
This need for computing power is particularly evident in training and inference for large, modern \ac{AI} models~\cite{shoeybi2020megatronlmtrainingmultibillionparameter, kaplan2020scalinglawsneurallanguage, Richards_2021}, and is driven by intense competition between both companies and countries in developing the best models and reaching \lq\lq\ac{AI} supremacy''~\cite{allison2020china,christie2021america,aisuperpower}.
Given the critical role of \acp{GPU} in this geopolitical competition, the United States has recently begun to implement controls on the export of the \acp{IC} designed and produced by U.S. companies on national security grounds -- to constrain advances in military and commercial systems fielded by potential adversary countries. 
In this section, we first provide an overview of recent export control regulations implemented by the United States. 
We then analyze the implications of these regulations on exports of the GPUs in our dataset, and the potential resulting performance gaps that denial might create.

\subsection{United States Export Control Regulations: Overview}
In October 2022, the \ac{BIS} within the United States Department of Commerce introduced the first regulation to control and restrict the export of high-performance AI chips to China~\cite{ic_bis_2022}. 
Specifically, the regulation added a new \ac{ECCN} 3A090, which imposed distinct conditions that a target advanced \ac{IC} had to meet simultaneously.\footnote{The regulation also includes ECCN 4A090, which controls computers, electronic assembles, and components containing \acp{IC} in \ac{ECCN} 3A090.}
These conditions can be summarized as follows: \begin{itemize}
    \item[] 1) Aggregate bidirectional \ac{IO} bandwidth $\ge$ 600 \ac{GBS}; and
    \item[] 2) \ac{TPP} $\ge$ 4800.
\end{itemize}
The \acf{TPP} metric (officially introduced in the following regulations) is defined as:
\begin{equation}
TPP = 2 \cdot MAC\_TOPS \cdot b 
\end{equation}
where $MAC\_TOPS$ is the number of theoretical peak \ac{TOPS} 
for \acf{MAC} computation, and $b$ is the bitwidth of the operation (\ac{TFLOPS} was used as a performance metric in the previous sections and is $2 \cdot MAC\_TOPS$ for \ac{FP} processing cores).
If an \ac{IC} can support various bit lengths that achieve different \ac{TPP} values (as in the case of GPU chips), the regulation restricts the highest \ac{TPP} value. Finally, the TPP values are for processing dense data; hence, we do not account for sparsity, even when supported.

Since this regulation prevented the export of GPUs such as A100 and H100 to China, NVIDIA launched derivative GPUs, namely the A800 and H800 (both in our dataset), to comply with the U.S. regulations while still serving the Chinese market. 
These GPUs incorporate chips that are basically identical to their non-export-controlled counterparts but reduce \ac{IO} bandwidth (A800 and H800) and/or FP64 performance (H800). 
A later 2023 revised regulation further tightened the restrictions and banned the export of these GPUs as well~\cite{ic_bis_2023}.
Similarly, another regulation from 2024~\cite{hbm_bis_1, hbm_bis_2} expanded the control scope to off-chip memories on GPU modules and included an additional constraint on \ac{HBM} density on exportable chips.

The latest regulation, announced in January 2025, is part of a “Framework for Artificial Intelligence Diffusion”~\cite{ic_bis}, further expanding existing export restrictions on high performance computing devices and HBM-based memory chips under \ac{ECCN} 3A090.
In particular, ECCN 3A090 features three distinct performance criteria:
\begin{itemize}
    \item 3A090.a controls \acp{IC} having either: \begin{itemize}
        \item[] 1) \ac{TPP} $\ge$ 4800; or
        \item[] 2) \ac{TPP} $\ge$ 1600 and performance density $\ge$ 5.92;
    \end{itemize}
    \item 3A090.b controls \acp{IC} having either: \begin{itemize}
    \item[] 1) 2400 $\le$ \ac{TPP} $<$ 4800 and 1.6 $\le$ performance density $<$ 5.92; or 
    \item[] 2) \ac{TPP} $\ge$ 1600 and 3.2 $\le$ performance density $<$ 5.92.
    \end{itemize}
    \item 3A090.c controls HBM-based memory chips having memory bandwidth density $>$ 2 GB/s per mm\textsuperscript{2}.
\end{itemize}
Performance density is defined as the ratio between TPP and the die area of a given AI chip, whereas memory bandwidth density is the memory bandwidth divided by the area of the memory package or stack.
It is important to note that advanced computing ICs that include co-packaged logic and HBM are not controlled by 3A090.c, even though they may be controlled by other ECCNs (e.g., 3A090.a or 3A090.b) based on their TPP and performance density.
Finally, ECCN 4A090 regulates the export of computers, electronic assemblies, and components that incorporate the aforementioned ICs.

Under the 2025 regulation, the top-performing GPUs that NVIDIA could export were the Ada Lovelace L2 and L20 and the Hopper H20, mainly designed for export purposes~\cite{semianalysis_h20}. However, after the U.S. approved the export of the H200 to ``approved customers" worldwide (on the condition that the U.S. collects a 25\% fee on the sales), this device has become the top-performing exportable NVIDIA GPU~\cite{h200_export, h200_export_2}.


\subsection{United States Export Control Regulations: Implications}
Having summarized evolving U.S. export control regulations, we can now analyze their potential impact on the GPU landscape. Specifically, we want to examine how such regulations affect the export of GPUs in our dataset and the potential performance gap that may arise between the United States and target countries.
Consequently, we evaluate the first regulation from 2022 through the last one from 2025. 
To this end, we employ the metrics available in our dataset, namely the peak computational performance (usually delivered by Tensor Cores), die size, and NVLink bandwidth, to evaluate the criteria based on TPP, performance density, and aggregated I/O bandwidth.
Lacking published public data on the die area of packaged HBM memory devices, we mainly ignore the parts of the 2024 and 2025 regulations covering HBM in our analysis, which would not prevent the export of advanced computing ICs anyway, as mentioned before.
Nonetheless, we note that exports of HBM chips conforming to the non-proprietary, 8-year-old international JEDEC standard HBM2 generally seem to be permitted, while chips meeting the newer JEDEC standards for the HBM2e and higher-performing standards (HBM3, HBM3e, HBM4) are generally restricted~\cite{morg24}. 

    

\begin{figure*}[t]
    \centering
    \begin{subfigure}{0.85\textwidth}
        \centering
        \includegraphics[width=\textwidth]{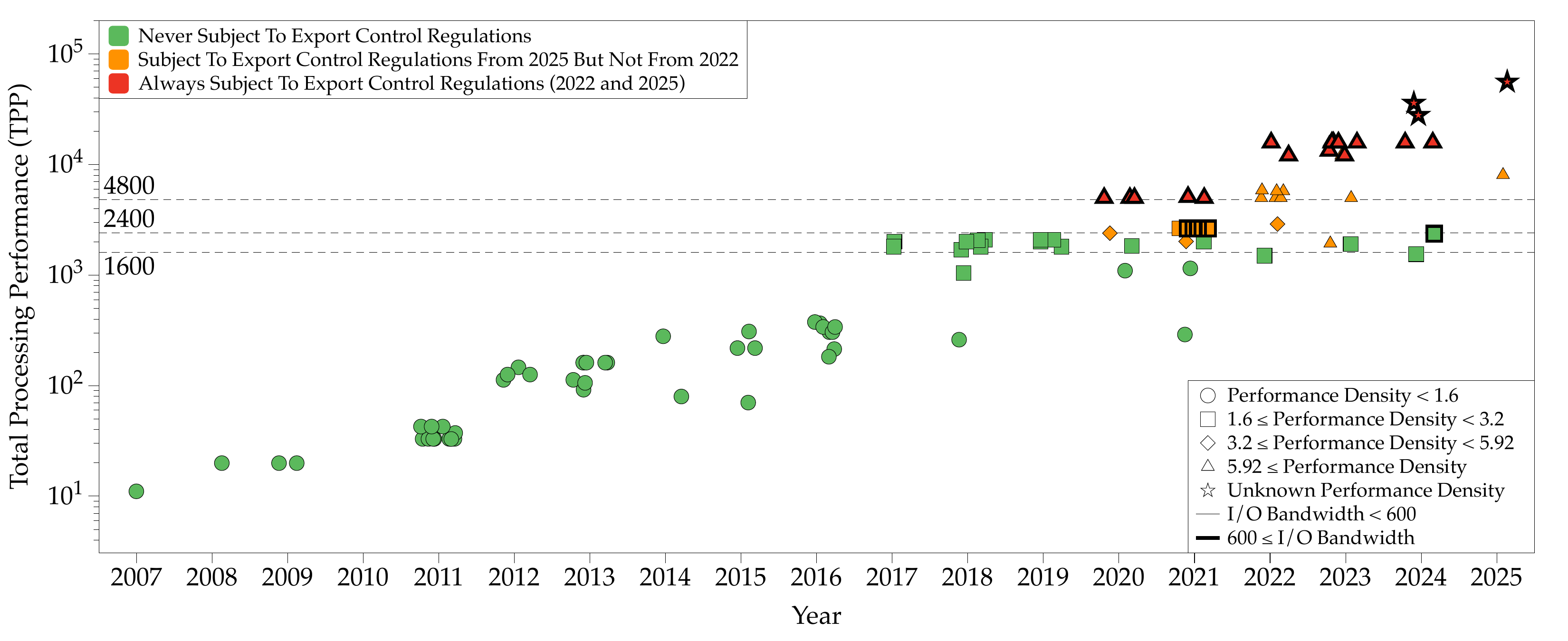}
        \caption{\acp{GPU} subject to export control regulations.}
        \label{fig:tpp_points}
    \end{subfigure}
    \begin{subfigure}{0.85\textwidth}
        \centering
        \includegraphics[width=\textwidth]{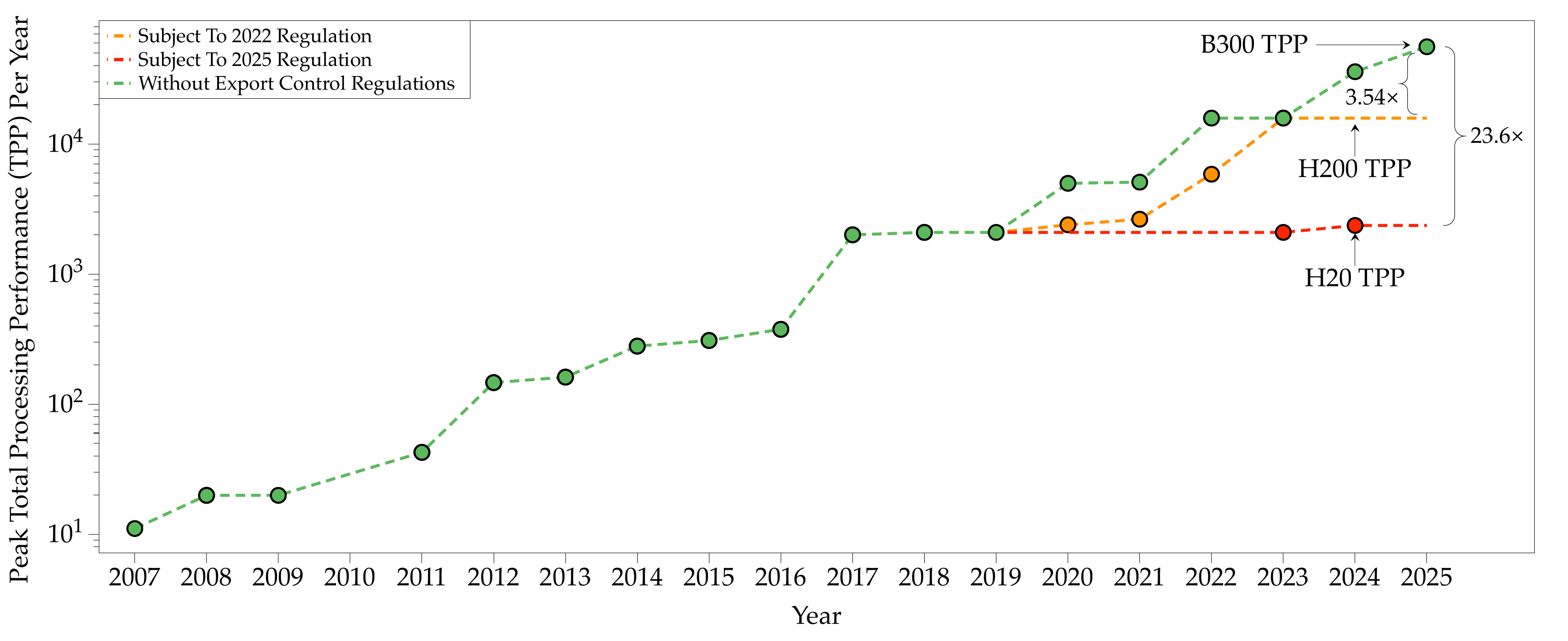}
        \caption{Peak exportable \acf{TPP} per year with and without export control regulations (we depict a non-decreasing pattern for each curve, meaning we keep the current maximum TPP over the years, even if the TPP in a subsequent year is lower).}
        \label{fig:tpp_peak}
    \end{subfigure}%
    \caption{Impact of the United States export control regulations from 2022 and 2025 on the \acp{GPU} in our dataset. We first show what \acp{GPU} are (or are not) subject to export control based on the regulations from 2022 and 2025. Then, we illustrate how the peak exportable \acf{TPP} per year changes when considering no export control regulations or the ones from 2022 and 2025 and the consequent performance gap. We apply horizontal jitter to the data points to prevent overplotting.}
    \label{fig:tpp_export_chart}
\end{figure*}

\Cref{fig:tpp_export_chart} shows the impact of the United States export control regulations in 2022 and 2025 through two charts. 
\Cref{fig:tpp_points} highlights how these regulations affect the export of the GPUs in our dataset and indicates the various \ac{TPP} (dotted lines), performance density (point shapes), and I/O (point border thickness) thresholds employed in the two target regulations.
Specifically, we can observe that no GPU released prior to 2019 is subject to either regulation, while most GPUs from 2020 and later are subject to export control regulations due to either the initial 2022 criteria or the more stringent criteria enacted in 2022 and 2025. 
\Cref{fig:tpp_peak} shows the consequences of such regulations, with lines indicating the peak TPP per year and how values change (or remain constant) under no export control regulations (green), the 2022 regulations (orange), and the 2025 regulations (red).
This Figure does not show trends but rather absolute/punctual peak TPP values per year, and marks the B300 as the GPU with the highest currently available TPP in the U.S. territory.
The outcome is a maximum 3.54$\times$ and 23.6$\times$ performance gap in computing power (measured as TPP) separating countries not subject to export control regulations from those subject to the 2022 and 2025 regulations, respectively, indicating how the 2025 regulations are much more restrictive than those from 2022.
If we account for the U.S. agreement on the export of the H200~\cite{h200_export, h200_export_2} in December 2025, the new potential gap drops to 3.54$\times$, which is interestingly equivalent to the gap we observed under the 2022 regulation.
Generally, the 2025 controls effectively restrict GPU performance to levels surpassed in 2020 (or in 2024, if we consider the recent policy update).
These controls do not account for other metrics such as off-chip memory size and bandwidth, which are particularly relevant in the AI field, the primary focus of these regulations.
Our estimations of the performance gap assume that the export controls are effective and that comparable GPU devices are not developed and sold by companies not subject to U.S. export controls. 

Another way to analyze the impact of U.S. export control regulations is to look at the potential gap in terms of TPP/\$. If we consider the B300, H20, and H200 GPUs (i.e., the GPUs with the highest TPP in the U.S., exportable under the 2025 regulations, and exportable after the recent policy update, respectively), we can observe that the initial gap between B300 and H20 in TPP/\$ was 5.16$\times$. However, this gap shrinks to 2.57$\times$ when replacing H20 with H200, implying that foreign countries can now potentially access 2$\times$ more TPP for the same investment.

Finally, despite this theoretical potential performance gap and consequent technological retardation, we should consider that these restrictions create powerful incentives to develop alternative technologies on the part of sanctioned countries, and stimulate creativity in both technological approaches, with researchers pursuing alternative innovative solutions and optimizations to keep pace with competitors, as well as creative ways to covertly bypass restrictions~\cite{hwang2022decoupling}. The recent release of AI models by the Chinese firm DeepSeek is an example of countering restrictions with innovation~\cite{guo2024deepseek, bi2024deepseek}; specifically, DeepSeek is reported to have relied on a few thousand H800 GPUs and implemented various optimization techniques to improve pipeline and data parallelism, counterbalancing the bandwidth limitations of the GPU utilized. 
Chinese memory producers are already producing their own HBM2 chips, have announced plans to ship HBM3 chips by 2026, and are rapidly narrowing their lag behind the US and Korean memory chip producers at the current leading edge of memory chip innovation~\cite{wang25,min25}. 
Furthermore, Chinese chip makers are currently producing homegrown GPU designs (such as Huawei's Ascend 910C, the best domestically produced Chinese AI chip~\cite{epoch2025whychinaisntabouttoleapaheadofthewestoncompute,cfr}) that, while not as performant as top NVIDIA designs, could close the gap in the future~\cite{wang25}. 
However, the current production capacity of Chinese companies like Huawei is a small fraction of NVIDIA's (5.3\% in 2025 and expected to drop to 2.2\% in 2027~\cite{cfr}); therefore, allowing China access to performant GPUs like the H200 (which is marginally more powerful than the Huawei Ascend 910C in computational performance and memory size/bandwidth) could increase China's ability to build and deploy large, cutting-edge AI models.
On top of that, we have to take into account that export controls do not prevent non-sanctioned users in a restricted destination country from accessing GPUs via offshore datacenter services in a non-restricted country.

Historically, export controls have met with limited long-run success in halting the eventual international diffusion of technology. 
The classic example is British export restrictions on textile machinery and knowhow during the first industrial revolution, which were wholly unsuccessful in preventing the diffusion of these technological innovations to rival industries in the United States and continental Europe, before they were abolished in the middle of the nineteenth century~\cite{har98,jer81,kel23}.
While undoubtedly making GPU compute acquisition more difficult and costly in the short term for geopolitical rivals, there is little current or historical evidence that suggests that any “performance gap” resulting from these controls is likely to persist for long. 
Economic history instead suggests that investment in continued innovation at the technological frontier is likely to be a more successful strategy to maintain technological leadership~\cite{abramovitz1986catching,mowery1999paths,mokyr2011gifts,mazzucato2013entrepreneurial,fleming2022breakthrough}. While lower cost strategies to quickly copy innovations and profitably follow a leader have at times been successful, it is difficult to see how technological leadership can be sustained without significant investments in frontier research and development. The current competitive dynamics of global R\&D investments in AI computing hardware would seem to support this thesis.

\section{Related Work and Limitations}\label{sec:limitations}

This section first reports similar relevant studies in the literature and compares them with our analysis. The section then examines the limitations of our work and suggests possible prospective research opportunities.

\subsection{Related Work}

The literature already contains studies and observations that aim to identify progress in GPUs (especially NVIDIA ones) from different perspectives. Probably the most widely-known example is Huang's law~\cite{8352557,huang_law,9623445,nvidia-keynote}, named after Jensen Huang, president and CEO of NVIDIA. This empirical law, which can be considered the counterpart of Moore's law for GPUs, has undergone several iterations as NVIDIA has released new GPUs and has reported AI compute gains over the years. Specifically, this law focuses on the peak theoretical performance each NVIDIA GPU microarchitecture can deliver for AI computations at the lowest supported data precision. If a given GPU does not support that specific precision (e.g., FP4), this law relies on performance at higher precisions (e.g., 8 or 16 bits). The iteration of Huang's law presented at COMPUTEX 2024~\cite{nvidia-keynote} spans from 2016 (19 TFLOPS FP16, Pascal P100) to 2024 (20,000 TFLOPS FP4 with sparsity, Blackwell B200) and reports a 1000$\times$ AI compute gain over 8 years. 

Huang's law provides a limited approach to GPU progress trends because it considers only the peak computing performance of 5 GPUs or microarchitectures~\cite{nvidia-keynote} (missing many years and devices) 
and combines different levels of data precision. 
Combining bitwidths in this way is questionable as a matter of historical progress, but is certainly incorrect if we want to predict future progress, since it is implausible from both physical and functional perspectives that bitwidths can continue to decrease.\footnote{Our work might seem to have a similar bias from GPUs before 2016, since we use their 32-bit performance, but this is because that is how those chips emulated 16-bit precision. We do not assume that bitwidths can continue declining.}
Consequently, Huang’s law yields an overestimated CAGR of 138\% (or DT of 0.8 years), which is 65 \acp{PPT} higher than the best CAGR we identified in our analysis (i.e., FP16 with sparsity for all datacenter GPUs). 
Unlike Huang's law, our work also provides a comprehensive analysis spanning a longer period, includes 102 NVIDIA GPUs, covers multiple key metrics, and is more consistent across different data-precision levels. 

Epoch AI has also done two studies, one in 2023 and one in 2024, analyzing progress in \ac{ML} hardware~\cite{epoch2023trendsinmachinelearninghardware, EpochMachineLearningHardware2024}. As shown in \Cref{table:comparison_epoch}, we find substantially different growth rates of progress than they do. These differences arise for two reasons: (1) Epoch AI’s analyses are much more heterogeneous, covering 170 devices. This approach gives those other devices more weight in the trend, whereas ours focuses on the standard and most-common architecture for ML and AI, NVIDIA datacenter GPUs.  (2) Epoch AI also chooses to group devices differently than we do. Specifically, they separate out traditional (e.g., CUDA) and tensor performance and analyze them separately. We instead pool these into a single analysis, because they compete for the same space on the chip and thus a decrease in one (e.g., FP64 CUDA Cores) can be used to get more of the other (Tensor Cores), as in the Blackwell Ultra B300.

Our analysis also considers important characteristics that Epoch AI misses, for instance, we account for sparsity in processor capabilities, measure FP64 performance improvement, and we track how computing performance is growing relative to off-chip memory bandwidth, which is crucial for understanding whether calculations will be bottlenecked by a ``memory wall''~\cite{10.1145/1555815.1555801}.

Finally, Baily et al. assessed the impact of generative AI on productivity and observed progress in prices and computing capabilities of NVIDIA and AMD GPUs, with a particular focus on the cost of compute (i.e., price per TFLOPS), reporting an annual decline by approximately 24 percent~\cite{baily2025generative,baily2026generative}. This result derives from the TechPowerUp GPU dataset~\cite{techpowerup-website}, which the authors employed as their source for price and computing performance. The TechPowerUp dataset includes a broad range of GPU classes (e.g., mobile, desktop, workstation, and datacenter) and primarily reports \ac{FP16}, \ac{FP32}, and \ac{FP64} performance values from CUDA Cores (for NVIDIA GPUs) or Streaming Processors (for AMD GPUs). Consequently, Baily et al. combine the cost of computing across different GPU classes, focus on \ac{FP32} performance only (based on the performance values reported in their Figure 8~\cite{baily2025generative}), and omit the contribution of Tensor Cores (or Matrix Cores on AMD GPUs). 
Furthermore, Baily et al. also report average annual rates for \ac{TFLOPS} growth (23\%) and price decline (1\%) on the considered GPU dataset~\cite{baily2025generative}.
These results largely differ from ours (i.e., growth rates of 51.4\% and 14.8\% for FP32 without sparsity and release price, respectively) due to the different scope of our research; indeed, we study a specific class of NVIDIA GPUs (i.e., the datacenter class), measure the \ac{TFLOPS}-per-price metric (rather than the price-per-TFLOPS one), analyze various data precisions separately, and account for the contribution of Tensor Cores to computing performance.

In summary, our work focuses on making apples-to-apples comparisons of the main processors used in AI by the dominant vendor of those processors. Within this set, we provide a comprehensive and consistent analysis spanning multiple metrics and configurations to assess various kinds of progress trends. Our work also includes an analysis of the potential impact of U.S. export control regulations on accessible GPU performance and the resultant capability gaps they may create.

\begin{table}[t]
    \centering
    \caption{Comparison of the yearly FP16 and FP32 improvements rates between our work and Epoch AI’s}
    \label{table:comparison_epoch}
    \resizebox{0.6\columnwidth}{!}{
    \begin{tabular}{lcccc}
    \toprule
    \multirow{2}{*}{\textbf{Metric}} & \multicolumn{2}{c}{\textbf{Progress [\%]}} & \multicolumn{2}{c}{\textbf{Progress / \$ [\%]}} \\
    & \textbf{Ours} & \textbf{Epoch AI} & \textbf{Ours} & \textbf{Epoch AI}\\
    \midrule
    16-bit Floating Point (w/o Sparsity) & 62 & 36 & 41 & 30 \\
    16-bit Floating Point (w/ Sparsity) & 73 & NA & 50 & NA \\
    32-bit Floating Point (w/o Sparsity) & 51 & 28 & 31 & 30 \\
    32-bit Floating Point (w/ sparsity) & 61 & NA & 40 & NA \\
    \bottomrule
    \end{tabular}
    }
\end{table}

\subsection{Limitations}
Although we believe this analysis offers an objective overview of the current pace of progress in NVIDIA GPUs, we believe there is still room for improvement. 
First, one of the most cumbersome parts of our analysis was GPU release price data collection. 
Top-notch datacenter GPUs are primarily purchased by corporate customers and authorized distributors, who often negotiate bulk deals that may include specific customizations. 
Typically, NVIDIA does not provide an official retail price in its announcements for such high-end devices, making the concept of a “release price at launch” more of a theoretical economic notion rather than an officially published number. 
Consequently, data collection was not straightforward or exhaustive, and we could not find precise information for all the GPUs reported in \Cref{table:collected_gpus} and their variants, leading us to drop roughly one-third of the GPUs from our per-dollar analysis (see \Cref{table:apx_data_stats} in \Cref{app:data_stats}).

Next, the discussion about the CAGRs and doubling times in \secref{sec:gpu_progress_analysis} is based solely on high-level metrics such as peak performance at different precisions. 
Effectively we are just observing the outcomes of enhancements occurring at the hardware level. 
A low-level analysis focusing on changes in the area and functionality of computing and on-chip memory components (i.e., caches) could unveil further insights about both past/current progress of GPUs and future potentially productive directions. 
Of course, such an analysis would require access to hardware design details that vendors do not tend to share publicly, as this could undermine their competitive advantage. 
While entities such as TechInsights, SemiAnalysis and Locuza publish articles~\cite{techinsights-article,semianalysis-article,locuza-article-1} that uncover the architectural specifications of some GPU chips, the limited number of GPUs examined hinders a more comprehensive analysis.

Another direction for analysis would be to distinguish between the physical design features of a given chip and the variants of those features actually sold in commercially available GPUs. In other words, even though the same chip design die model (same die size and transistor count) is used across multiple GPU boards, each device sold typically implements only a subset of the full theoretical capabilities fabricated in the die model. For instance, the GH100 GPU chip includes 18432 FP32 CUDA Cores in its full implementation~\cite{h100}. However, all the Hopper GPU boards reported in this study, based on the same chip, have either 16896 or 14592 FP32 CUDA Cores, depending on the form factor. The difference between capabilities in actual models sold and those available theoretically in the chip hardware design is undoubtedly due, to some extent, to design redundancy intended to correct for fabrication defects. By having “spare” CUDA cores available, defective cores can be fused off and replaced with spares, thus increasing the yield rate of good dies on a processed silicon wafer and reducing fabrication cost. This aspect is crucial for NVIDIA GPUs, which are among the largest chips ever manufactured, often as large as the current state of chip fabrication technology permits. Constant defect densities in semiconductor fabrication imply that larger chips will, on average, have more defects. Redundant features turn otherwise defective chips from scrap into saleable products. 

Finally, as mentioned previously, our analysis utilizes peak values for performance and bandwidth, and TDP for power consumption. Please note that other literature studies employ the same approach based on theoretical peak values for computational performance and off-chip memory bandwidth, as well as TDP~\cite{huang_law,epoch2023trendsinmachinelearninghardware,EpochMachineLearningHardware2024}. Although this choice permits analysis of ideal/theoretical characteristics of these GPUs, we recognize that further discussion based on real-world benchmarks 
would be beneficial for understanding actual workload performance based on the workload type and on the role of software optimizations. Moreover, the theoretical computing performance values reported in NVIDIA documentation are based on the boost clock, which can only be used for a limited time, as it would quickly push the GPU past its TDP and raise the chip's operating temperature to alarming levels, causing a fast down-clocking.
Similarly, TDP is generally around two-thirds of the maximum power draw, meaning it does not represent either peak or average power consumed during a real-world computation workload~\cite{10.5555/3207796}. 
Given the extensive analysis required, we plan to pursue additional research employing real benchmark values to more accurately portray trade-offs in design choices addressing different computing application domains (e.g., AI).

\section{Conclusion}\label{sec:conclusion}


GPUs have reshaped and are continuing to reshape the fields of computer science and computer architecture. What was initially a supporting device designed to offload graphics processing from the CPU has undoubtedly become even more critical and compelling than the CPU itself, especially for AI. Indeed, NVIDIA GPUs were one of the main drivers of the current AI revolution, and, even though there may be other devices that are more suitable and specialized for AI workloads, such as Google's TPUs, they remain the state-of-the-art approach for training and inference of AI models, thanks to both their hardware features and software infrastructure. On the hardware side, NVIDIA has applied several improvements and introduced new capabilities to its GPUs over the years. Components like Tensor Cores or, in general, specialized units/accelerators for AI basic computations were then adopted and implemented similarly by other CPU and GPU vendors for their devices, such as \acp{NPU} in Intel CPUs and Matrix Cores in AMD GPUs. The role of GPUs (and AI processors, in general) has been so prominent recently that the U.S. has introduced national security regulations to control their export.

As the demand for computational power continues to grow at a pace never before seen, examining both past and current technical progress is crucial for identifying future limitations in scientific research. For this reason, our work analyzes significant trends in NVIDIA GPUs. Based on our curated dataset of datacenter GPUs released since the mid-2000s, we provide insights about the technical, per-memory bandwidth, per-dollar, and per-watt CAGRs and DTs, foreshadowing expectations for the near future.\footnote{We plan to release this dataset as part of the MIT Processor Database: \url{https://processordb.mit.edu}.}
We summarize the key insights from the analysis presented in this paper.

\ourparagraph{Insight 1 - GPU progress surpassed Moore's Law to keep pace with AI growth} Moore's Law has long been the primary indicator of progress in semiconductor technology, predicting that the complexity of ICs would double every two years~\cite{moore1975progress}, with consequent implications on performance growth.
While the CPU improvement rate is slowing down~\cite{7878935}, the progress of GPUs and, generally, AI hardware is outpacing Moore's Law prediction, as our analysis of past and current trends in NVIDIA datacenter GPUs indicates; for instance, FP16 performance doubles every 1.27 to 1.43 years (with and without sparsity, respectively).
In general, the main driver of GPU and specialized hardware progress is the enormous demand for computing required by AI, whose performance and memory requirements are growing at unprecedented rates (e.g., doubling times of 6 to 10 months for computing demand~\cite{epoch2025trainingcomputedecomposition, Sevilla_2022} and 30$\times$ per year for longest context windows~\cite{epoch2025contextwindows}), pushing the production cycles of the semiconductor industry. As a consequence, NVIDIA is implementing architectural trade-offs to provide customers with fast turnaround times for training increasingly large AI models, while deprioritizing features such as high-precision performance (e.g., FP64).

\ourparagraph{Insight 2 - chip-level enhancements boosted GPU performance} A significant contribution to the progress of NVIDIA GPUs derives from improvements at the chip level. Features such as FP16 hardware support, specialized AI units (e.g., Tensor Cores), and increasingly large single- and multi-die GPUs have certainly paid off as a short-term solution to boost performance and keep pace with increasing computing demand. The future will tell us more about the sustainability of such an approach and its diminishing returns.


\ourparagraph{Insight 3 - the computing power of FP16 and FP32 doubles in less than 2 years} Our analysis across our entire dataset indicates the doubling times for FP16 and FP32 performance are 1.43y and 1.67y, respectively. As mentioned in the previous insight, this result is a direct consequence of the performance boost yielded by the Tensor Cores since 2017 (for FP16) and 2020 (for FP32). Moreover, the estimated DTs for FP16 and FP32 decrease further when sparsity support is accounted for (1.27y and 1.46y, respectively), although sparsity is not always applicable.

\ourparagraph{Insight 4 - FP64 compute units are becoming a lower-priority resource} The growth in FP64 performance is not as prominent as the one of FP16 and FP32. NVIDIA usually sacrifices FP64 compute capability in desktop/workstation-oriented GPUs, limiting the number of FP64 units or omitting FP64 support in the Tensor Cores. This decision implies a slower growth rate, as we observed in the configuration with fewer FP64 CUDA Cores per SM (20.1\% across the entire dataset). Similarly, even if pure datacenter GPUs used to provide more FP64 units and Tensor Core support (though not as prominent as for other data precisions), their progress trend will probably decrease in the future, as even top-notch GPUs such as the Blackwell Ultra B300 feature limited FP64 computing capabilities, presumably because they are mainly designed for the AI field, which does not require greater precision~\cite{dettmers2022llmint88bitmatrixmultiplication, gupta2015deeplearninglimitednumerical, courbariaux2015trainingdeepneuralnetworks} (the same may be true also for other scientific fields~\cite{domke2019double}), making NVIDIA repurpose those transistors to other data precisions.

\ourparagraph{Insight 5 - HBM boosted off-chip memory size and bandwidth} Off-chip memory characteristics leaped forward with the introduction of HBM technology. While it took HBM longer to outpace GDDR in terms of memory size, it immediately contributed to the growth of off-chip memory bandwidth, making HBM an essential component of high-end NVIDIA GPUs. Specifically, HBM alone yielded 31.8\% and 26.8\% CAGRs for off-chip memory size and bandwidth, which are 13.7 and 14.1 \acp{PPT} higher than GDDR alone.

\ourparagraph{Insight 6 - the growth in release price and TDP for best-in-class GPUs is almost 2$\times$ that of the datacenter lineup} Release price and, especially, TDP are slower than other metrics in terms of growth, which is a positive outcome. However, the subset of top-performing GPUs per year yields higher growth rates for both release price and TDP (22.7\% and 8.6\%, respectively) than NVIDIA's overall datacenter GPU lineup (14.8\% and 4.6\%), suggesting that best-in-class GPUs are becoming increasingly expensive and power-hungry.
Supporting evidence for the second point is that GPU power demand now represents approximately 40\% of total power usage in AI data centers~\cite{epoch2025gpuspowerusageinaidatacenters}.

\ourparagraph{Insight 7 - computing power grows faster than off-chip memory bandwidth} The technical improvements per bandwidth tell an interesting story. Both the subset of top-performing GPUs per year and the entire dataset yield generally favorable progress trends for FP16 and FP32 (though not as high as the related performance growth rates): 30.2\% and 32.5\% for FP16 and 21.8\% and 23.5\% for FP32, respectively. These trends imply that the growth of FP16 and FP32 metrics outpaces off-chip memory bandwidth, potentially making it a bottleneck for computations that heavily depend on off-chip data transfers and creating the so-called memory or bandwidth wall. Interestingly, the FP64 technical improvements per bandwidth show a different picture, especially for GPUs with lower FP64 capabilities, whose growth rate is almost 0\%, indicating a fairly balanced growth in both FP64 performance and off-chip memory bandwidth.
In general, one way to both address the memory wall and increase computational performance is to reduce the data precision, because the same amount of bits/bytes would contain more data, and the compute units would be potentially able to perform more operations per clock cycle (assuming hardware support for a given data precision and proportional performance scaling). On the other hand, such an approach introduces accuracy loss as the bitwidth decreases, also reaching the physical limit set by the number of bits. Furthermore, low data precision requires quantization algorithms that introduce additional overhead and are usually data- and domain-specific.
Finally, another critical factor likely to widen the gap between computing and off-chip memory capabilities is the current shortage of memory chips; indeed, the ever-increasing demand for off-chip memory in AI data centers is creating a supply-and-demand imbalance, leading to higher memory chip prices and a potential slowdown in AI development~\cite{chip_supply_1, chip_supply_2}.

\ourparagraph{Insight 8 - the technical improvement per watt grows faster than the per-dollar one} 
In terms of technical improvement per dollar and per watt, we observe that the increase in release prices results in less technical improvement per dollar than per watt, as the release price metric grows faster over time than the TDP one. In particular, our analysis shows that TFLOPS/W (compute per variable energy cost) increases by 1.3$\times$ to 2.9$\times$ relative to TFLOPS/\$ (compute capacity per fixed capital cost). 
Meeting the demand for the ever more powerful hardware needed to sustain modern workloads should shorten the lifetimes of the most powerful, cutting-edge computing devices, particularly AI-oriented ones. It would be useful to confirm that lifecycle compute costs are increasingly dominated by fixed investments in hardware purchases rather than by variable costs of energy (electricity), as implied by the contrasting trends in price and energy use per unit of theoretical computing output that we have documented in this paper.

\ourparagraph{Insight 9 - AMD and Intel are narrowing the gap, but not enough to affect NVIDIA’s dominance}
Thanks to its architectural innovations and the resulting computational performance, NVIDIA secured its role as the dominant GPU vendor in both the desktop and datacenter domains and as the main solution for AI workloads in the mid-2010s. Even though the other GPU vendors, particularly AMD, managed to catch up with NVIDIA GPUs through similar architectural enhancements and, in recent years, even surpassed NVIDIA’s performance (e.g., FP16 and FP64), the improvement margin remains too thin to significantly affect the market or NVIDIA’s role and justify a shift toward other vendors. On top of that, we need to remember that a theoretically powerful piece of hardware requires a solid software stack to be usable in an easy and effective way. While AMD and Intel have invested heavily in ROCm and oneAPI over the last few years, NVIDIA's software stack (i.e., CUDA and its libraries) remains ahead of the competition as the most efficient, widespread, and supported solution for GPU (and AI) computing.

\ourparagraph{Insight 10 - the recent update to export control regulations reduced the performance gap from 23.6$\times$ to 3.54$\times$} Finally, our analysis of recent U.S. export control regulations on advanced AI chips identifies a theoretical, potential peak-performance gap, measured in TPP, of 23.6$\times$ that might open up between the U.S. and restricted countries if these export controls had their intended effect.
However, as the U.S. recently agreed to allow NVIDIA to export the H200 GPUs, this performance gap dropped to 3.54$\times$. While the H200 provides marginal gains over Chinese domestic devices (e.g., Huawei Ascend 910C) in terms of computational performance and memory size/bandwidth, this agreement could enable China to improve its ability to build large cutting-edge AI models significantly, thanks to the access to a greater number of GPUs than its domestic production would provide. Allegedly, Chinese companies have already ordered more than 2 million H200 for 2026~\cite{h200_demand}.

\bigskip

In conclusion, while our work provides a retrospective analysis of the progress trends of NVIDIA datacenter GPUs released to date, one could question whether the identified trends can be projected into the future. Despite the new GPUs announced by NVIDIA for the following years promise to deliver unprecedented performance~\cite{gtc_2025}, we are doubtful that the aforementioned trends will continue at the rates discussed in this paper. The primary rationale for such uncertainty is that many improvements come from short-term solutions that may yield diminishing returns soon (e.g., reducing data precision or multi-die GPUs). For this reason, we plan to discuss the future prospects and the barriers these trends will face in a forthcoming article.

\begin{acks}
The Authors would like to thank Filippo Carloni, Andrew Lohn, Albert Reuther, Larry Rudolph, and Ana Trišović for their valuable feedback and kind support.
\end{acks}

\bibliographystyle{ACM-Reference-Format}
\bibliography{references}

%
\clearpage
\acresetall
\appendix
\renewcommand{\thefigure}{A\arabic{figure}}
\renewcommand{\thetable}{A\arabic{table}}
\setcounter{figure}{0}
\setcounter{table}{0}
\crefalias{section}{appendix}



\section{Collected GPU Metrics}\label{app:collected_metrics}

\begin{table*}[ht]
    \caption{Collected GPU metrics/features per microarchitecture (in bold the metrics we employed in the NVIDIA datacenter GPU analysis, vendor comparison, or export control analysis)}
    \label{table:apx_gpu_metrics}
    \begin{minipage}{\textwidth}
    \centering
    \resizebox{\textwidth}{!}{
    \begin{tabular}{lcccccccccccc}
    \toprule
    \multirow{2}{*}{Metric/Feature} & \multirow{2}{*}{Tesla} & \multirow{2}{*}{Fermi} & \multirow{2}{*}{Kepler} & \multirow{2}{*}{Maxwell} & \multirow{2}{*}{Pascal} & \multirow{2}{*}{Volta} & \multirow{2}{*}{Turing} & \multirow{2}{*}{Ampere} & Ada & \multirow{2}{*}{Hopper} & \multirow{2}{*}{Blackwell} & Blackwell\\
    &  &  &  &  &  &  &  &  & Lovelace &  &  & Ultra\\
    \midrule
\textbf{Chip Name} & \cmark & \cmark & \cmark & \cmark & \cmark & \cmark & \cmark & \cmark & \cmark & \cmark & \cmark & \cmark \\
\textbf{Release Year} & \cmark & \cmark & \cmark & \cmark & \cmark & \cmark & \cmark & \cmark & \cmark & \cmark & \cmark & \cmark \\
\textbf{Launch Price} & \myhalfcheck & \myhalfcheck & \myhalfcheck & \myhalfcheck & \myhalfcheck & \myhalfcheck & \myhalfcheck & \myhalfcheck & \myhalfcheck & \myhalfcheck & \cmark & \cmark \\
Manufacturing Process & \cmark & \cmark & \cmark & \cmark & \cmark & \cmark & \cmark & \cmark & \cmark & \cmark & \cmark & \cmark \\
Transistor Count & \cmark & \cmark & \cmark & \cmark & \cmark & \cmark & \cmark & \cmark & \cmark & \cmark & \cmark & \cmark \\
\textbf{Die Size} & \cmark & \cmark & \cmark & \cmark & \cmark & \cmark & \cmark & \cmark & \cmark & \cmark & \myhalfcheck & \xmark \\
Die Density & \cmark & \cmark & \cmark & \cmark & \cmark & \cmark & \cmark & \cmark & \cmark & \cmark & \myhalfcheck & \xmark \\
Bus Interface & \cmark & \cmark & \cmark & \cmark & \cmark & \cmark & \cmark & \cmark & \cmark & \cmark & \cmark & \cmark \\
Form Factor & \cmark & \cmark & \cmark & \cmark & \cmark & \cmark & \cmark & \cmark & \cmark & \cmark & \cmark & \cmark \\
\textbf{Thermal Design Power} & \cmark & \cmark & \cmark & \cmark & \cmark & \cmark & \cmark & \cmark & \cmark & \cmark & \cmark & \cmark \\
Base Clock & \cmark & \cmark & \cmark & \cmark & \cmark & \cmark & \cmark & \cmark & \myhalfcheck & \myhalfcheck & \myhalfcheck & \xmark \\
Boost Clock & \cmark & \cmark & \myhalfcheck & \cmark & \cmark & \cmark & \cmark & \cmark & \myhalfcheck & \myhalfcheck & \cmark & \cmark \\
\acf{SM} & \cmark & \cmark & \cmark & \cmark & \cmark & \cmark & \cmark & \cmark & \myhalfcheck & \myhalfcheck & \cmark & \cmark \\
\acfp{PB} per \ac{SM} & \cmark & \cmark & \cmark & \cmark & \cmark & \cmark & \cmark & \cmark & \cmark & \cmark & \cmark & \cmark \\
CUDA Cores: INT32 per \ac{PB} & \cmark & \cmark & \cmark & \cmark & \cmark & \cmark & \cmark & \cmark & \cmark & \cmark & \cmark & \cmark \\
CUDA Cores: INT32 per \ac{SM} & \cmark & \cmark & \cmark & \cmark & \cmark & \cmark & \cmark & \cmark & \cmark & \cmark & \cmark & \cmark \\
CUDA Cores: INT32 per GPU & \cmark & \cmark & \cmark & \cmark & \cmark & \cmark & \cmark & \cmark & \myhalfcheck & \myhalfcheck & \cmark & \cmark \\
CUDA Cores: FP32 per \ac{PB} & \cmark & \cmark & \cmark & \cmark & \cmark & \cmark & \cmark & \cmark & \cmark & \cmark & \cmark & \cmark \\
CUDA Cores: FP32 per \ac{SM} & \cmark & \cmark & \cmark & \cmark & \cmark & \cmark & \cmark & \cmark & \cmark & \cmark & \cmark & \cmark \\
CUDA Cores: FP32 per GPU & \cmark & \cmark & \cmark & \cmark & \cmark & \cmark & \cmark & \cmark & \myhalfcheck & \myhalfcheck & \cmark & \cmark \\
CUDA Cores: FP64 per \ac{PB} & \myhalfcheck & \cmark & \cmark & \cmark & \cmark & \cmark & \myna & \myhalfcheck & \myna & \myhalfcheck & \myhalfcheck & \myna \\
\textbf{CUDA Cores: FP64 per \ac{SM}} & \myhalfcheck & \cmark & \cmark & \cmark & \cmark & \cmark & \cmark & \cmark & \cmark & \cmark & \cmark & \cmark \\
CUDA Cores: FP64 per GPU & \myhalfcheck & \cmark & \cmark & \cmark & \cmark & \cmark & \cmark & \cmark & \myhalfcheck & \myhalfcheck & \cmark & \cmark \\
\textbf{CUDA Cores: Peak FP16 Performance} & \myna & \myna & \myna & \myna & \cmark & \cmark & \cmark & \cmark & \cmark & \cmark & \cmark & \cmark\\
CUDA Cores: Peak \ac{BF16} Performance & \myna & \myna & \myna & \myna & \myna & \myna & \myna & \cmark & \cmark & \cmark & \cmark & \cmark\\
CUDA Cores: Peak INT32 Performance & \cmark & \cmark & \cmark & \cmark & \cmark & \cmark & \cmark & \cmark & \cmark & \cmark & \myhalfcheck & \myna \\
\textbf{CUDA Cores: Peak FP32 Performance} & \cmark & \cmark & \cmark & \cmark & \cmark & \cmark & \cmark & \cmark & \cmark & \cmark & \cmark & \cmark \\
\textbf{CUDA Cores: Peak FP64 Performance} & \myhalfcheck & \cmark & \cmark & \cmark & \cmark & \cmark & \cmark & \cmark & \myhalfcheck & \cmark & \cmark & \cmark \\
Tensor Cores: Boost Clock & \myna  & \myna  & \myna  & \myna  & \myna  & \myna  & \myna  & \myna  & \myna & \cmark & \myna & \myna\\
Tensor Cores: Units per \ac{PB} & \myna & \myna & \myna & \myna & \myna & \cmark & \cmark & \cmark & \cmark & \cmark & \cmark & \cmark\\
Tensor Cores: Units per \ac{SM} & \myna & \myna & \myna & \myna & \myna & \cmark & \cmark & \cmark & \cmark & \cmark & \cmark & \cmark\\
Tensor Cores: Units per GPU & \myna & \myna & \myna & \myna & \myna & \cmark & \cmark & \cmark & \myhalfcheck & \myhalfcheck & \cmark & \cmark\\
Tensor Cores: Generation & \myna & \myna & \myna & \myna & \myna & \cmark & \cmark & \cmark & \cmark & \cmark & \cmark & \cmark\\

Tensor Cores: FP16 \ac{MAC} per Cycle & \myna & \myna & \myna & \myna & \myna & \cmark & \cmark & \cmark & \cmark & \cmark & \cmark & \cmark\\
Tensor Cores: Supported Data Types & \myna & \myna & \myna & \myna & \myna & \cmark & \cmark & \cmark & \cmark & \cmark & \cmark & \cmark\\
\textbf{Tensor Cores: Sparsity Support} & \myna & \myna & \myna & \myna & \myna & \myna & \myna & \cmark & \cmark & \cmark & \cmark & \cmark\\
\textbf{Tensor Cores: Peak INT4 Performance} & \myna & \myna & \myna & \myna & \myna & \myna & \cmark & \cmark & \cmark & \xmark & \myhalfcheck & \xmark \\
\textbf{Tensor Cores: Peak FP4 Performance} & \myna & \myna & \myna & \myna & \myna & \myna & \myna & \myna & \myna & \myna & \cmark & \cmark \\
\textbf{Tensor Cores: Peak FP6 Performance} & \myna & \myna & \myna & \myna & \myna & \myna & \myna & \myna & \myna & \myna & \cmark & \cmark \\
\textbf{Tensor Cores: Peak INT8 Performance} & \myna & \myna & \myna & \myna & \myna & \myna & \cmark & \cmark & \cmark & \cmark & \cmark & \cmark\\
\textbf{Tensor Cores: Peak FP8 Performance} & \myna & \myna & \myna & \myna & \myna & \myna & \myna & \myna & \cmark & \cmark & \cmark & \cmark\\
\textbf{Tensor Cores: Peak FP16 Performance} & \myna & \myna & \myna & \myna & \myna & \cmark & \cmark & \cmark & \cmark & \cmark & \cmark & \cmark\\
\textbf{Tensor Cores: Peak \ac{BF16} Performance} & \myna & \myna & \myna & \myna & \myna & \myna & \myna & \cmark & \cmark & \cmark & \cmark & \cmark\\
\textbf{Tensor Cores: Peak \ac{TF32} Performance} & \myna & \myna & \myna & \myna & \myna & \myna & \myna & \cmark & \cmark & \cmark & \cmark & \cmark\\
\textbf{Tensor Cores: Peak FP64 Performance} & \myna & \myna & \myna & \myna & \myna & \myna & \myna & \myhalfcheck & \myna & \cmark & \myhalfcheck & \cmark\\
Register File Size per \ac{SM} & \xmark & \cmark & \cmark & \cmark & \cmark & \cmark & \cmark & \cmark & \cmark & \cmark & \cmark & \cmark \\
Register File Size per GPU & \xmark & \cmark & \cmark & \cmark & \cmark & \cmark & \cmark & \cmark & \myhalfcheck & \myhalfcheck & \cmark & \cmark \\
Shared Memory Size per \ac{SM} & \cmark & \cmark & \cmark & \cmark & \cmark & \cmark & \cmark & \cmark & \cmark & \cmark & \cmark & \cmark \\
L1 Cache Size per \ac{SM} & \xmark & \cmark & \cmark & \cmark & \cmark & \cmark & \cmark & \cmark & \cmark & \cmark & \cmark & \cmark \\
L1 Cache Size per GPU & \cmark & \cmark & \cmark & \cmark & \cmark & \cmark & \cmark & \cmark & \myhalfcheck & \myhalfcheck & \cmark & \cmark \\
L2 Cache Size per GPU & \cmark & \cmark & \cmark & \cmark & \cmark & \cmark & \cmark & \cmark & \cmark & \cmark & \myhalfcheck & \xmark \\
\textbf{Memory Type} & \cmark & \cmark & \cmark & \cmark & \cmark & \cmark & \cmark & \cmark & \cmark & \cmark & \cmark & \cmark \\
\textbf{Memory Size} & \cmark & \cmark & \cmark & \cmark & \cmark & \cmark & \cmark & \cmark & \cmark & \cmark & \cmark & \cmark \\
Memory Bus & \cmark & \cmark & \cmark & \cmark & \cmark & \cmark & \cmark & \cmark & \cmark & \cmark & \cmark & \cmark \\
\textbf{Memory Bandwidth} & \cmark & \cmark & \cmark & \cmark & \cmark & \cmark & \cmark & \cmark & \cmark & \cmark & \cmark & \cmark \\
Memory Data Rate & \cmark & \cmark & \cmark & \cmark & \cmark & \cmark & \cmark & \cmark & \cmark & \cmark & \cmark & \cmark \\
\textbf{NVLink Bandwidth} & \myna & \myna & \myna & \myna & \myhalfcheck & \myhalfcheck & \myna & \myhalfcheck & \myna & \cmark & \myhalfcheck & \cmark \\
\bottomrule
    \end{tabular}
    }
    \footnotesize{\cmark: metric/feature collected for all GPUs belonging to a given microarchitecture \\ \myhalfcheck: metric/feature collected for some GPUs belonging to a given microarchitecture (e.g., metric not available/implemented or retrievable for some GPUs)\\ \xmark: metric/feature technically available/implemented but not retrievable for any GPU belonging to a given microarchitecture \\ \myna: metric/feature not available/implemented for a given microarchitecture}.
\end{minipage}
\end{table*}

\clearpage

\section{Data Statistics}\label{app:data_stats}

\begin{table*}[ht]
 \caption{Statistics on the data used to fit the exponential models on the technical improvements per metric and their per-memory bandwidth, per-dollar, and per-watt ratios for top-performing GPUs and all datacenter GPUs}
 \label{table:apx_data_stats}
 \centering
 \resizebox{\textwidth}{!}{
 \begin{tabular}{lcccccccccccccc}
 \toprule
 \multirow{3}{*}{\textbf{Metric}} &\multirow{3}{*}{\textbf{Configuration}} & \multirow{3}{*}{\textbf{Unit}} & \multicolumn{6}{c}{\textbf{Top-Performing GPUs}} & \multicolumn{6}{c}{\textbf{All Datacenter GPUs}}\\
 \cmidrule(lr){4-9} \cmidrule(lr){10-15}
 & & & \textbf{Sample} & \multirow{2}{*}{\textbf{Mean}} & \multirow{2}{*}{\textbf{Std}} & \multirow{2}{*}{\textbf{Min}} & \multirow{2}{*}{\textbf{Max}} & \multirow{2}{*}{\textbf{R\textsuperscript{2}}} & \textbf{Sample} & \multirow{2}{*}{\textbf{Mean}} & \multirow{2}{*}{\textbf{Std}} & \multirow{2}{*}{\textbf{Min}} & \multirow{2}{*}{\textbf{Max}} & \multirow{2}{*}{\textbf{R\textsuperscript{2}}}\\
 & & & \textbf{Size}  & & & & & & \textbf{Size} \\
 \addlinespace
 \midrule
 \addlinespace
 \multicolumn{15}{c}{\textbf{Technical Improvement}}\\
 \addlinespace
 \midrule
 \addlinespace
\multirow{2}{*}{16-bit Floating Point} & w/o Sparsity & \multirow{2}{*}{TFLOPS} & 18 & 410.47 & 728.29 & 0.35 & 2250 & 0.98 & 102 & 253.60 & 432.94 & 0.35 & 2250 & 0.91 \\
 & w/ Sparsity & & - & - & - & - & - & - & 102 & 488.06 & 873.42 & 0.35 & 4500 & 0.93 \\
 \addlinespace
 \arrayrulecolor{black!15}\cmidrule[0.1em]{2-15}
 \addlinespace
 \multirow{2}{*}{32-bit Floating Point} & w/o Sparsity & \multirow{2}{*}{TFLOPS} & 18 & 198.14 & 367.39 & 0.35 & 1125 & 0.96 & 102 & 119.93 & 218.26 & 0.35 & 1125 & 0.88 \\
 & w/ Sparsity & & - & - & - & - & - & - & 102 & 235.92 & 438.52 & 0.35 & 2250 & 0.88 \\
 \addlinespace
 \arrayrulecolor{black!15}\cmidrule[0.1em]{2-15}
 \addlinespace
 \multirow{2}{*}{64-bit Floating Point} & $>$ 8 FP64 CUDA Cores per SM & \multirow{2}{*}{TFLOPS} & 12 & 19.87 & 22.95 & 0.52 & 66.91 & 0.94 & 66 & 16.14 & 21.21 & 0.51 & 66.91 & 0.94 \\
 & $\leq$ 8 FP64 CUDA Cores per SM & & - & - & - & - & - & - & 33 & 0.60 & 0.52 & 6.86E-02 & 1.97 & 0.69 \\
 \addlinespace
 \arrayrulecolor{black!15}\cmidrule[0.1em]{2-15}
 \addlinespace
 \multirow{3}{*}{Off-chip Memory Size} & GDDR and HBM & \multirow{3}{*}{GB} & 18 & 53.64 & 72.63 & 1.50 & 270 & 0.96 & 102 & 38.69 & 45.01 & 1.50 & 270 & 0.81 \\
 & GDDR Only & & - & - & - & - & - & - & 55 & 17.50 & 17.56 & 1.50 & 96 & 0.75 \\
 & HBM Only & & - & - & - & - & - & - & 47 & 63.49 & 54.05 & 12 & 270 & 0.76 \\
 \addlinespace
 \arrayrulecolor{black!15}\cmidrule[0.1em]{2-15}
 \addlinespace
 \multirow{3}{*}{Off-chip Memory Bandwidth} & GDDR and HBM & \multirow{3}{*}{GB/s} & 18 & 1791.80 & 2423.33 & 76.80 & 7700 & 0.98 & 102 & 1203.62 & 1551.97 & 76.80 & 7700 & 0.73 \\
 & GDDR Only & & - & - & - & - & - & - & 55 & 366.89 & 289.29 & 76.80 & 1597 & 0.68 \\
 & HBM Only & & - & - & - & - & - & - & 47 & 2182.78 & 1836.72 & 549.10 & 7700 & 0.76 \\
 \addlinespace
 \arrayrulecolor{black!15}\cmidrule[0.1em]{2-15}
 \addlinespace
 Release Price & - & \$ & 18 & 16781.34 & 16409.75 & 1300 & 55000 & 0.97 & 68 & 12253.58 & 11807.77 & 1300 & 55000 & 0.54 \\
 \addlinespace
 \arrayrulecolor{black!15}\cmidrule[0.1em]{2-15}
 \addlinespace
 Thermal Design Power & - & W & 18 & 383.05 & 270.49 & 170.90 & 1100 & 0.73 & 102 & 300.51 & 183.70 & 60 & 1100 & 0.16 \\
 \addlinespace
 \arrayrulecolor{black}\midrule
 \addlinespace

 \multicolumn{15}{c}{\textbf{Technical Improvement Per Memory Bandwidth}}\\
 \addlinespace
 \midrule
 \addlinespace
 \multirow{2}{*}{16-bit Floating Point} & w/o Sparsity & \multirow{2}{*}{TFLOPS / GB/s} & 18 & 0.11 & 0.11 & 4.50E-03 & 0.30 & 0.94 & 102 & 0.13 & 0.12 & 4.50E-03 & 0.49 & 0.89 \\
 & w/ Sparsity & & - & - & - & - & - & - & 102 & 0.24 & 0.26 & 4.50E-03 & 0.98 & 0.92 \\
 \addlinespace
 \arrayrulecolor{black!15}\cmidrule[0.1em]{2-15}
 \addlinespace
 \multirow{2}{*}{32-bit Floating Point} & w/o Sparsity & \multirow{2}{*}{TFLOPS / GB/s} & 18 & 4.82E-02 & 5.35E-02 & 4.50E-03 & 0.15 & 0.86 & 102 & 6.04E-02 & 6.12E-02 & 4.50E-03 & 0.24 & 0.78 \\
 & w/ Sparsity & & - & - & - & - & - & - & 102 & 0.11 & 0.13 & 4.50E-03 & 0.49 & 0.79 \\
 \addlinespace
 \arrayrulecolor{black!15}\cmidrule[0.1em]{2-15}
 \addlinespace
 \multirow{2}{*}{64-bit Floating Point} & $>$ 8 FP64 CUDA Cores per SM & \multirow{2}{*}{TFLOPS / GB/s} & 12 & 8.92E-03 & 4.59E-03 & 3.59E-03 & 2.00E-02 & 0.38 & 66 & 9.13E-03 & 5.92E-03 & 3.29E-03 & 3.31E-02 & 0.66 \\
 & $\leq$ 8 FP64 CUDA Cores per SM &  & - & - & - & - & - & - & 33 & 9.00E-04 & 4.50E-04 & 1.62E-04 & 1.66E-03 & 0.00 \\
 \addlinespace
 \arrayrulecolor{black}\midrule
 \addlinespace
 \multicolumn{15}{c}{\textbf{Technical Improvement Per Dollar}}\\
 \addlinespace
 \midrule
 \addlinespace
 \multirow{2}{*}{16-bit Floating Point} & w/o Sparsity & \multirow{2}{*}{TFLOPS / \$} & 18 & 1.16E-02 & 1.50E-02 & 2.31E-04 & 5.00E-02 & 0.93 & 68 & 1.14E-02 & 1.35E-02 & 1.88E-04 & 5.00E-02 & 0.90 \\
 & w/ Sparsity & & - & - & - & - & - & - & 68 & 2.13E-02 & 2.77E-02 & 1.88E-04 & 0.10 & 0.92 \\
 \addlinespace
 \arrayrulecolor{black!15}\cmidrule[0.1em]{2-15}
 \addlinespace
 \multirow{2}{*}{32-bit Floating Point} & w/o Sparsity & \multirow{2}{*}{TFLOPS / \$} & 18 & 5.31E-03 & 7.60E-03 & 2.31E-04 & 2.50E-02 & 0.87 & 68 & 5.32E-03 & 6.52E-03 & 1.88E-04 & 2.50E-02 & 0.87 \\
 & w/ Sparsity & & - & - & - & - & - & - & 68 & 1.01E-02 & 1.34E-02 & 1.88E-04 & 5.00E-02 & 0.87 \\
 \addlinespace
 \arrayrulecolor{black!15}\cmidrule[0.1em]{2-15}
 \addlinespace
 \multirow{2}{*}{64-bit Floating Point} & $>$ 8 FP64 CUDA Cores per SM & \multirow{2}{*}{TFLOPS / \$} & 12 & 7.95E-04 & 5.25E-04 & 1.29E-04 & 1.99E-03 & 0.73 & 43 & 7.20E-04 & 5.73E-04 & 9.39E-05 & 2.23E-03 & 0.82 \\
 & $\leq$ 8 FP64 CUDA Cores per SM & & - & - & - & - & - & - & 23 & 6.79E-05 & 7.00E-05 & 1.98E-05 & 3.49E-04 & 0.12 \\
 \addlinespace
 \arrayrulecolor{black!15}\cmidrule[0.1em]{2-15}
 \addlinespace
 \multirow{3}{*}{Off-chip Memory Size} & GDDR and HBM & \multirow{3}{*}{GB / \$} & 18 & 2.53E-03 & 1.26E-03 & 1.00E-03 & 4.91E-03 & 0.42 & 68 & 3.07E-03 & 2.27E-03 & 7.79E-04 & 1.16E-02 & 0.46 \\
 & GDDR Only & & - & - & - & - & - & - & 41 & 6.29E-02 & 4.20E-02 & 2.67E-02 & 0.21 & 0.17 \\
 & HBM Only & & - & - & - & - & - & - & 27 & 0.12 & 5.69E-02 & 6.11E-02 & 0.33 & 0.29 \\
 \addlinespace
 \arrayrulecolor{black!15}\cmidrule[0.1em]{2-15}
 \addlinespace
 \multirow{3}{*}{Off-chip Memory Bandwidth} & GDDR and HBM & \multirow{3}{*}{GB/s / \$} & 18 & 8.12E-02 & 3.53E-02 & 3.56E-02 & 0.17 & 0.49 & 68 & 8.51E-02 & 5.54E-02 & 2.67E-02 & 0.33 & 0.40 \\
 & GDDR Only & & - & - & - & - & - & - & 41 & 6.29E-02 & 4.20E-02 & 2.67E-02 & 0.21 & 0.17 \\
 & HBM Only & & - & - & - & - & - & - & 27 & 0.12 & 5.69E-02 & 6.11E-02 & 0.33 & 0.29 \\
 \addlinespace
 \arrayrulecolor{black}\midrule
 \addlinespace
 \multicolumn{15}{c}{\textbf{Technical Improvement Per Watt}}\\
 \addlinespace
 \arrayrulecolor{black}\midrule
 \addlinespace
 \multirow{2}{*}{16-bit Floating Point} & w/o Sparsity & \multirow{2}{*}{TFLOPS / W} & 18 & 0.62 & 0.81 & 2.02E-03 & 2.25 & 0.96 & 102 & 0.60 & 0.65 & 2.02E-03 & 2.50 & 0.92 \\
 & w/ Sparsity & & - & - & - & - & - & - & 102 & 1.12 & 1.34 & 2.02E-03 & 5.00 & 0.94 \\
 \addlinespace
 \arrayrulecolor{black!15}\cmidrule[0.1em]{2-15}
 \addlinespace
 \multirow{2}{*}{32-bit Floating Point} & w/o Sparsity & \multirow{2}{*}{TFLOPS / W} & 18 & 0.29 & 0.41 & 2.02E-03 & 1.12 & 0.96 & 102 & 0.28 & 0.33 & 2.02E-03 & 1.25 & 0.91 \\
 & w/ Sparsity & & - & - & - & - & - & - & 102 & 0.53 & 0.68 & 2.02E-03 & 2.50 & 0.90 \\ 
 \addlinespace
 \arrayrulecolor{black!15}\cmidrule[0.1em]{2-15}
 \addlinespace
 \multirow{2}{*}{64-bit Floating Point} & $>$ 8 FP64 CUDA Cores per SM & \multirow{2}{*}{TFLOPS / W} & 12 & 4.34E-02 & 4.30E-02 & 2.16E-03 & 0.15 & 0.85 & 66 & 4.02E-02 & 3.93E-02 & 2.09E-03 & 0.15 & 0.93 \\
 & $\leq$ 8 FP64 CUDA Cores per SM & & - & - & - & - & - & - & 33 & 2.50E-03 & 1.89E-03 & 4.14E-04 & 6.57E-03 & 0.58 \\
 \addlinespace
 \arrayrulecolor{black!15}\cmidrule[0.1em]{2-15}
 \addlinespace
 \multirow{3}{*}{Off-chip Memory Size} & GDDR and HBM & \multirow{3}{*}{GB / W} & 18 & 0.10 & 8.67E-02 & 8.78E-03 & 0.27 & 0.90 & 102 & 0.12 & 8.06E-02 & 8.78E-03 & 0.33 & 0.80 \\
 & GDDR Only & & - & - & - & - & - & - & 55 & 8.46E-02 & 7.55E-02 & 8.78E-03 & 0.33 & 0.77 \\
 & HBM Only & & - & - & - & - & - & - & 47 & 0.15 & 7.11E-02 & 4.80E-02 & 0.32 & 0.61 \\
 \addlinespace
 \arrayrulecolor{black!15}\cmidrule[0.1em]{2-15}
 \addlinespace
 \multirow{3}{*}{Off-chip Memory Bandwidth} & GDDR and HBM & \multirow{3}{*}{GB/s / W} & 18 & 3.42 & 2.89 & 0.45 & 9.85 & 0.94 & 102 & 3.36 & 2.37 & 0.45 & 11 & 0.80 \\
 & GDDR Only & & - & - & - & - & - & - & 55 & 1.75 & 1.20 & 0.45 & 4.57 & 0.76 \\
 & HBM Only & & - & - & - & - & - & - & 47 & 5.24 & 1.99 & 2.20 & 11 & 0.66 \\
 \addlinespace
 \arrayrulecolor{black}\bottomrule
 \addlinespace
 \end{tabular}
 }
\end{table*}










%
%









\end{document}